\documentclass[11pt,a4paper]{article}
\pdfoutput=1
\usepackage{jheppub}

\usepackage{amsmath,amssymb}
\usepackage{xifthen}
\usepackage{graphicx}
\graphicspath{{figures-symmetric/},{figures/}}
\usepackage{color}
\usepackage[utf8]{inputenc}

\newcommand{\figref}[1]{figure~\ref{#1}}
\newcommand{\Figref}[1]{Figure~\ref{#1}}
\newcommand{\secref}[1]{section~\ref{#1}}

\newcommand{\appref}[1]{appendix~\ref{#1}}

\newcommand{\op}[1]{\ensuremath{\hat{#1}}}
\newcommand{\abs}[1]{\ensuremath{\left|#1\right|}}
\newcommand{\tr}[1]{\ensuremath{\mathrm{Tr}\left(#1\right)}}

\newcommand{\ud}[2]{\ensuremath{\mathrm{d}^{#2}#1\,}}
\newcommand{\dd}{\ensuremath{\mathrm{d}}}
\newcommand{\qwf}{\hat{\Psi}}

\newcommand{\pert}[1]{\ensuremath{\delta#1}}

\newcommand{\hamD}{\ensuremath{\mathcal{H}}}
\newcommand{\lagDdel}[1][]{\ensuremath{\mathcal{L}_{\delta}^{\ifthenelse{\isempty{#1}}{}{(#1)}}}}
\newcommand{\eomPhiT}{\ensuremath{\mathcal{E}}}

\newcommand{\ham}{\ensuremath{H}}
\newcommand{\wf}{\ensuremath{\psi}}

\newcommand{\floq}{\ensuremath{\Lambda}}
\newcommand{\floqMax}{\ensuremath{\floq_{\rm max}}}

\newcommand{\gS}{\ensuremath{g}}
\newcommand{\gC}{\ensuremath{g_c}}
\newcommand{\nuT}{\ensuremath{\nu}}
\newcommand{\nuZero}{\ensuremath{\nu_0}}
\newcommand{\nuAmp}{\ensuremath{\delta}}
\newcommand{\nuFreq}{\ensuremath{\omega}}

\newcommand{\lPar}{\ensuremath{\lambda}}
\newcommand{\chemP}{\mu}

\newcommand{\rhoP}{\ensuremath{\varrho}}
\newcommand{\rhoN}{\ensuremath{\varepsilon}}
\newcommand{\phiP}{\ensuremath{\vartheta}}
\newcommand{\phiN}{\ensuremath{\varphi}}
\newcommand{\rhoPAve}{\ensuremath{\bar{n}}}
\newcommand{\rhoNAve}{\ensuremath{\bar{\rhoN}}}
\newcommand{\phiPAve}{\ensuremath{\bar{\phiP}}}
\newcommand{\phiNAve}{\ensuremath{\bar{\phiN}}}

\newcommand{\chemPBG}{\ensuremath{\mu}}
\newcommand{\potBG}{\ensuremath{U_{\rm bg}}}
\newcommand{\rhoBG}{\ensuremath{\rhoPAve}}
\newcommand{\dRhoBG}{\ensuremath{\rhoNAve}}
\newcommand{\dphiBG}{\ensuremath{\phiNAve}}
\newcommand{\tphiBG}{\ensuremath{\phiPAve}}
\newcommand{\drhoRat}{\ensuremath{x}}

\newcommand{\eig}{\ensuremath{\eta}}
\newcommand{\nuBG}{\ensuremath{\bar{\nu}_{\rm bg}}}

\newcommand{\rhoM}{\ensuremath{n}}                
\newcommand{\rhoPert}{\ensuremath{\pert{\rho}}}    
\newcommand{\rhoNorm}{\ensuremath{\delta}}        
\newcommand{\phaseM}{\ensuremath{\bar{\phi}}}     
\newcommand{\phasePert}{\ensuremath{\delta\phi}}  
\newcommand{\delRatTot}{\ensuremath{\pert{\xi}}}  
\newcommand{\delRatDiff}{\ensuremath{\pert{\zeta}}}
\newcommand{\delPhTot}{\ensuremath{\pert{\vartheta}}}
\newcommand{\delPhDiff}{\ensuremath{\pert{\varphi}}}

\newcommand{\dphaseD}{\ensuremath{\pert{\phiN}}}     
\newcommand{\drhoDK}{\ensuremath{\delta\tilde{\zeta}_{\bf k}}}
\newcommand{\dphaseDK}{\ensuremath{\delta\tilde{\phiN}_{\bf k}}}

\newcommand{\tG}{\ensuremath{\tilde{t}}} 
\newcommand{\xG}{\ensuremath{\tilde{x}}}  
\newcommand{\tRel}{\ensuremath{\tilde{t}_{s}}}  
\newcommand{\xRel}{\ensuremath{\tilde{x}_{s}}}  

\newcommand{\kBarSym}{\ensuremath{\tilde{k}_{\rm s}}}
\newcommand{\nuBarSym}{\ensuremath{\tilde{\nu}_{\rm s}}}

\newcommand{\wBarSym}{\ensuremath{\tilde{\nuFreq}_s}}
\newcommand{\csTot}{\ensuremath{c_{\rm tot}}}
\newcommand{\csRel}{\ensuremath{c_{\rm rel}}}
\newcommand{\keffsq}{\ensuremath{\kappa^2}} 
\newcommand{\keffsqt}{\ensuremath{\tilde{\kappa}^2}}
\newcommand{\kFloq}{\ensuremath{k_{\rm floq}}}
\newcommand{\kBarFloq}{\ensuremath{\tilde{k}_{\rm floq}}}
\newcommand{\noDim}[1]{\ensuremath{\tilde{#1}}}
\newcommand{\cphi}{\ensuremath{\sigma}}  

\author[a]{Jonathan Braden}
\author[b,c]{Matthew C.\ Johnson}
\author[a,d]{Hiranya V.\ Peiris}
\author[e,f]{Silke Weinfurtner}

\affiliation[a]{Department of Physics and Astronomy, University College London, Gower Street, London, WC1E 6BT, UK}
\affiliation[b]{Department of Physics and Astronomy, York University, Toronto, Ontario, M3J 1P3, Canada}
\affiliation[c]{Perimeter Institute for Theoretical Physics, 31 Caroline St. N, Waterloo, ON, N2L 2Y5, Canada}
\affiliation[d]{Oskar Klein Center for Cosmoparticle Physics, Stockholm University, AlbaNova, Stockholm, SE-106 91, Sweden}
\affiliation[e]{School of Mathematical Sciences, University of Nottingham, University Park, Nottingham, NG7 2RD, UK}
\affiliation[f]{Centre for the Mathematics and Theoretical Physics of Quantum Non-Equilibrium Systems, University of Nottingham, Nottingham NG7 2RD, UK}

\emailAdd{j.braden@ucl.ac.uk}
\emailAdd{mjohnson@perimeterinstitute.ca}
\emailAdd{h.peiris@ucl.ac.uk}
\emailAdd{silke.weinfurtner@nottingham.ac.uk}

\abstract{
Analog condensed matter systems present an exciting opportunity to simulate early Universe models in table-top experiments.
We consider a recent proposal for an analog condensed matter experiment to simulate the relativistic quantum decay of the false vacuum.
In the proposed experiment, two ultra-cold condensates are coupled via a time-varying radio-frequency field.
The relative phase of the two condensates in this system is approximately described by a relativistic scalar field with a potential possessing a series of false and true vacuum local minima.
If the system is set up in a false vacuum, it would then decay to a true vacuum via quantum mechanical tunnelling.
Should such an experiment be realized, it would be possible to answer a number of open questions regarding non-perturbative phenomena in quantum field theory and early Universe cosmology.
In this paper, we illustrate a possible obstruction: the time-varying coupling that is invoked to create a false vacuum for the long-wavelength modes of the condensate leads to a destabilization of shorter wavelength modes within the system via parametric resonance. We focus on an idealized setup in which the two condensates have identical properties and identical background densities.
Describing the system by the coupled Gross-Pitaevskii equations (GPE), we use the machinery of Floquet theory to perform a linear stability analysis, calculating the wavenumber associated with the first instability band for a variety of experimental parameters.
However, we demonstrate that, by tuning the frequency of the time-varying coupling, it may be possible to push the first instability band outside the validity of the GPE, where dissipative effects are expected to damp any instabilities.
This provides a viable range of experimental parameters to perform analog experiments of false vacuum decay.
}
\title{Towards the cold atom analog false vacuum}

\begin{document}
\maketitle

\section{Introduction}
False vacuum minima are ubiquitous in relativistic field theories with complex interaction potentials.
Although classically stable, large spatial regions where the fields are trapped in such false vacua can undergo decay via the nucleation of bubbles (either through quantum~\cite{Coleman:1977py,Callan:1977pt} or classical statistical fluctuations~\cite{Linde:1981zj}).
In cosmology, such transitions are a crucial ingredient in many incarnations of the multiverse (see e.g.~\cite{Aguirre:2007gy} for a review), and may have also played a role in the early history of our own Hubble volume if symmetry breaking phase transitions occurred at some sufficiently high temperature~\cite{Linde:2005ht}.
Since cosmological evolution is intrinsically time-dependent, it is of great interest to have a time-dependent understanding of the phenomenology of bubble nucleation.
Unfortunately, standard descriptions of this process~\cite{Coleman:1977py,Callan:1977pt} cannot properly tackle the time-dependent nature of the problem.
Instead, the temporal dependence is introduced in an ad hoc manner with bubble nucleation events assumed to be independent and distributed in the spacetime according to a Poisson process.
While some analytic progress has been made in understanding a Lorentzian picture of quantum tunnelling~\cite{Turok:2013dfa,Feldbrugge:2017kzv}, here we focus on a possible laboratory experiment that can be used as a quantum computer for simulating vacuum decay in ensembles of "Universes" in order to understand the true quantum dynamics of first order phase transitions. Should such experiments be feasible, they would provide insight into fundamental open questions such as:
\begin{itemize}
\item In the standard instanton description of bubble nucleation~\cite{Coleman:1977py,Callan:1977pt}, one imagines that a spherically symmetric bubble with a critical radius ``appears'' via quantum tunneling, and eventually is well-described by the classical field configuration of a bubble of true vacuum expanding with a constant acceleration. What precisely happens before and during the ``appearance'' of the bubble? Are there any precursors to bubble nucleation, corresponding to intermediate field configurations, before a final transition to the expanding bubble? How should we understand the transition to the classical limit at late times?
\item Instanton techniques do not give a complete description of multi-bubble configurations, and nucleation events are considered to be completely independent. In the full dynamics, are there non-trivial correlations between nucleation events? If so, there are potentially important implications for if/how percolation occurs in quantum first order phase transitions in quantum field theory. 
\item A deeper understanding of percolation could also inform our understanding of eternal inflation, where accelerated expansion of space can prevent percolation. Are there important correlations that give rise to clusters of bubbles, with different properties than the clusters that appear for uncorrelated nucleation events~\cite{Guth:1981uk,Guth:1982pn}? If so, there are potentially observable consequences from the collisions between bubbles in such clusters~\cite{Aguirre:2007an,Chang:2007eq,Feeney_etal:2010jj,Wainwright:2013lea}.
\item Since the bubble interior has the symmetries of an open Friedman-Robertson-Walker (FRW) spacetime at late times, bubble nucleation represents the quantum creation of a Universe. Are there any novel observable signatures of the quantum origins of the Universe stored in the resulting field configurations?
\end{itemize}

Recently, Fialko et al~\cite{Fialko:2014xba,Fialko:2016ggg} proposed such a condensed matter system to simulate the false vacuum decay of a relativistic scalar field in the laboratory.
This is one amongst a variety of analog condensed matter systems that have been proposed to study various elements of relativistic quantum field theory.
An interesting aspect of the false vacuum decay proposal is that it purports to simulate fully nonlinear quantum dynamics of an interacting field theory.
This is in contrast to many proposed analog gravity systems (such as those involving analog Hawking and Unruh radiation~\cite{Weinfurtner:2010nu,Euve:2015vml,Steinhauer:2015saa}, superradiance~\cite{Torres:2016iee}, and cosmological particle production~\cite{Jaskula:2012ab,Eckel:2017uqx}), which consider the kinematics of small amplitude non-interacting fluctuations propagating on an externally controlled classical background.
Similarly, proposals to study the dynamics of solitons~\cite{1984lcb..book.....N} and nonlinear structures such as oscillons (see e.g.~\cite{Copeland:1995fq,Amin:2011hj}) typically focus on classical evolution in a nonlinear system~\cite{PhysRevA.91.023631}.
In the linearized cases, the connection to relativistic fields is made through a recreation of the linear equations of motion for the fluctuations, rather than through nonlinear phenomena induced by mode-mode coupling.

In its simplest form, the proposed system for false vacuum decay consists of a pair of ultra cold gases with 2$\to$2 intra- and inter-particle scattering interactions.
The two particle species are then coupled through a radio-frequency field, which leads to a bilinear coupling describing the direct conversion of one species into the other. 
Below the critical temperature, the atoms condense into a Bose-Einstein condensate (BEC) phase of long-wavelength collective motion, whose dynamics can be modeled by the coupled Gross-Pitaevskii equations (GPEs); see e.g.~\cite{2013EJPh...34..247R} for a pedagogical review.
This allows for a quantitative treatment of the dynamics of the long wavelength mean field condensate.

The (complex) condensates are conveniently rewritten in terms of the squares of their moduli (interpreted as a local condensate particle density) and their phases (whose gradients are proportional to the local condensate velocities).
By integrating out fluctuations in the number density, the effective dynamics of the two phase variables is approximated by a pair of relativistic scalar fields.
In the limit that the bilinear coupling goes to zero, the Hamiltonian is independent of the phase of each condensate, resulting in two global $U(1)$ symmetries.
The introduction of a non-zero coupling breaks the symmetry associated with a global change to the relative phase, while preserving the symmetry under a simultaneous rotation of both condensates.
This generates a sinusoidal potential for the relative phase, while the total phase remains massless.

The final step to obtain an analog false vacuum is to modulate the bilinear coupling periodically in time.
For sufficiently long wavelengths, the effect of this temporal modulation is to convert the local maxima of the original sinusoidal potential into a series of local minima, thus leading to an effective scalar field with a series of degenerate false vacua.
However, the same oscillating coupling that is introduced to stabilize the zero mode of the system simultaneously induces additional instabilities at higher wavenumbers that are not accounted for in the time-averaged effective potential. This is an example of parametric resonance, which has been well-studied in the context of early Universe cosmology~\cite{Kofman:1997yn,Kofman:1994rk,Traschen:1990sw} (see~\cite{Amin:2014eta} for a recent review), and has been experimentally verified in a single-component BEC~\cite{Jaskula:2012ab}.
If these wavenumbers are part of the long-wavelength collective modes described by the coupled GPEs, then the exponential growth of these modes will eventually destabilize the condensate as a whole, simultaneously invalidating the conclusion that the field is sitting in a false vacuum.
Although we have described a particular implementation, the key feature for false vacuum decay is that the GPEs describe the dynamics of the long-wavelength condensates, but not the specific details of the atom-atom interactions at very short wavelengths.

In this paper, we perform a detailed linear stability analysis of the proposed analog false vacuum setup, assuming the initial state can be described as a homogeneous background with small inhomogeneous perturbations on top.
We assume the GPE provides an adequate description of the condensate dynamics, and perform the calculation in the full ``UV''-theory on the condensed matter side, rather than in the effective scalar description.
We demonstrate the stabilization of the zero-mode within linear perturbation theory as we increase the amplitude of the modulation of the bilinear coupling.
For modes below the healing length, we also show that the growth rates and transition to linear stability agree well with the predictions of the effective action for the effective relativistic scalar field.
As anticipated above, while the zero-mode is stabilized through the increasing amplitude of the modulations, we simultaneously observe the emergence of exponentially unstable modes with nonzero wavenumbers.

We identify the oscillation frequency of the modulations as the key parameter, with the characteristic wavenumber of the resonant instabilities scaling approximately as the square root of the driving frequency on scales shorter than the healing length.
If these exponentially unstable modes are correctly modeled by the GPE, they will lead to a catastrophic fragmentation of the proposed false vacuum.
For sufficiently short wavelengths we expect viscous and stochastic corrections to the GPE will appear.  By pushing the unstable modes into the viscous regime, we thus expect they will damp away, thus preserving the metastability of the system.
However, the required frequency must not be too large, or additional degrees of freedom (e.g.\ radial modes for condensates in a ring trap, or mediating transitions for hyperfine splitting) can be excited.
Therefore, to fully assess the feasibility of studying false vacuum decay in the lab, the corrections to the GPE at short wavelengths must be modeled, and the corresponding impact of these corrections on the dynamically induced instabilities must be understood.
We do not undertake this comprehensive analysis here.
However, the theoretical link between the GPE and a relativistic scalar undergoing false vacuum decay is interesting in its own right, and allows for a new theoretical toolbox to be applied to the problem.
Therefore, even absent a complete understanding of the experimental feasibility, theoretical investigation of the proposed analog system is of interest.

In order to consider the simplest physical situation, and avoid additional technical complications, in this paper we assume that the two condensates have equal effective masses, equal self-interaction scattering rates, and equal background densities (which we refer to as the symmetric limit).
This symmetric setup illustrates the basic mechanisms at play for more general choices of condensate parameters.
Of course, in a real experiment these symmetry assumptions may not be realized, so it is also of interest to perturb off the symmetric limit (while still maintaining the assumption of a homogeneous background).
The linearized analysis in these asymmetric cases will be presented in a forthcoming publication, where we describe the rich additional dynamics that emerge.

The remainder of this paper is organised as follows.
In~\secref{sec:bec-review} we briefly review the physics of dilute gas Bose-Einstein condensates (BECs), present the equations of motion describing the condensates, and briefly comment on the regime of validity of the equations.
We then connect the BEC dynamics to relativistic scalar field theory in~\secref{sec:effective-action}, and explicitly demonstrate how a false vacuum effective potential can be engineered for the relativistic scalar degree of freedom.
Next, we obtain suitable homogeneous background solutions around which to perform our fluctuation analysis in~\secref{sec:homogeneous-bg}.
The main results of the paper are contained in~\secref{sec:linear-symmetric}, where we analyze the linear fluctuation dynamics around the proposed false vacuum using the machinery of Floquet theory.
Finally, we conclude with a summary and some future directions in~\secref{sec:conclusion}.
To streamline the main presentation, technical details regarding the emergence of a relativistic scalar field (\appref{app:effective-action}) and the derivation of the linearized perturbation equations (\appref{app:linear}) are contained in a pair of appendices.

\section{Coupled Bose Condensates}\label{sec:bec-review}
The condensed matter system under consideration is a 2-component BEC, an ultra-cold gas of $N$ atoms containing two single-particle states $\vert 1 \rangle$ and $\vert 2 \rangle$ such that $N_1+N_2=N$. An example for such a system is $^{87}\mathrm{Rb}$ in two different hyperfine states. If the gas is ultra-cold and highly diluted, the atom-atom interaction potential can be approximated by an $S$-wave contact potential. With this approximation, there are three kinds of atom-atom contact interactions present in the system. They are parameterized by three coupling constants: $g_{11}$ and $g_{22}$ within each component, and $g_{12}=g_{21}$ across components. Each hyperfine state has a different total angular momentum, and hence slightly different energies. One can employ external fields to set up three-level scattering processes to drive transitions between the hyperfine states with transition rate $\nuT$.
\subsection{Model Hamiltonian and Gross-Pitaevskii Equations}
We first derive the relevant equations for the 2-component BEC.
The boson field annihilation $\qwf_i$ and creation operators $\qwf^\dag_i$ for the single-particle states satisfy
\begin{align}
\left[ \qwf_i ({\bf x}), \qwf_j ({\bf y})\right] &= \left[ \qwf^\dag_i ({\bf x}), \qwf_j^\dag ({\bf y})\right] = 0 \\
\left[ \qwf_i ({\bf x}), \qwf^\dag_j ({\bf y})\right] &= \delta_{ij} \delta({\bf x-y}) \, .
\end{align}
In terms of these, the Hamiltonian for the BECs in external trapping potentials $V_{{\rm ext},i}(x)$ is given by $\op{\ham} = \int\ud{x}{d}\op{\hamD}$, with the Hamiltonian density
\begin{eqnarray}
  \label{eqn:ham}
  \hat\hamD =
 - \qwf^\dag_i \frac{\hbar^2\nabla^2}{2 m_i} \qwf_i
 + \qwf^\dag_i V_{{\rm ext},i} \qwf_i
 + \frac{g_{ij}}{2} \qwf_i^\dag \qwf_j^\dag \qwf_i \qwf_j - \frac{\nu}{2}\sigma^{\rm x}_{ij}\qwf_i^\dag \qwf_j \, .
\end{eqnarray}
where
\begin{equation}
  \sigma^{\rm x} =
  \left(\begin{array}{cc}0&1\\1&0\end{array}\right) \, .
\end{equation}
The coupled Gross-Pitaevskii equations for the 2-component BEC follow from the Heisenberg equations of motion,
\begin{equation}
  \label{eqn:gpe}
  i\hbar\frac{\partial\qwf_i}{\partial t} = \left[-\frac{\hbar^2}{2m_i}\nabla^2 + V_{{\rm ext},i}(x) + g_{ij}\qwf_j^\dag\qwf_j\right]\qwf_i - \nu \sigma_{ij}^{\rm x}\qwf_j \, .
\end{equation}
When working with the path integral in the next section, it will be convenient to have an explicitly real Hamiltonian density.
Integrating by parts, we can re-express~\eqref{eqn:ham} as
\begin{equation}
  \label{eqn:ham-dens-real}
   \hat\hamD =
   \frac{\hbar^2}{2 m_i}\nabla\qwf_i^\dag\cdot\nabla\qwf_i
     + \qwf^\dag_i V_{{\rm ext},i} \qwf_i
     + \frac{g_{ij}}{2} \qwf_i^\dag \qwf_j^\dag \qwf_i \qwf_j - \frac{\nu}{2}\sigma_{ij}^{\rm x}\qwf_i^\dag \qwf_j
\end{equation}
up to a boundary term.
In our analysis, we will ignore the effects of the external potentials $V_{{\rm ext},i}$.
We could imagine, for example, that the condensate is trapped in a ring with strong confining potentials in the two orthogonal directions; see~\cite{Eckel:2017uqx} for such a setup.
The transverse excitations can then be integrated out leaving a theory for the longitudinal modes with no effective trapping potential. However, one needs to be careful not to excite radial eigenmodes as exhibited in~\cite{Eckel:2017uqx}.
Setting $V_{{\rm eff},i} = 0$, it is then consistent to consider dynamics around a homogeneous background.
Including the external potential results in a spatially inhomogeneous ground state that must be perturbed.
This significantly complicates the linear stability analysis carried out here, so we leave investigation of this case to future work.

If cooled down to sufficiently low temperatures the system undergoes spontaneous symmetry breaking through the formation of a background condensate.
It is convenient to redefine our field operators 
\begin{equation}
\qwf_i= \sqrt{\hat{\rho}_i} \, e^{i\hat{\phi}_i} \, .
\end{equation}
The Gross-Pitaevskii equations are now obtained by replacing the quantum operators $\qwf_i$ in~\eqref{eqn:gpe} with
\begin{equation}
  \label{eqn:psi-bg-def}
  \psi_i = \sqrt{\rho_i}e^{i\phi_i} \, .
\end{equation}
This obviously neglects nonlinear corrections from interactions amongst the quantum fluctuations around $\psi_i$.

Through direct transformation of the resulting equations we obtain
\begin{subequations}\label{eqn:eom-rho-phi-species}
\begin{align}
  \hbar\dot{\rho}_i &= -\frac{\hbar^2}{m_i}\left(\rho_i\nabla^2\phi_i + \nabla \rho_i\cdot\nabla\phi_i\right) - 2\nu\sum_{j\neq i}\sqrt{\rho_i\rho_j}\sin(\phi_j-\phi_i) \\
  \hbar\dot{\phi}_i &= \frac{\hbar^2}{2m_i}\left(\frac{\nabla^2\rho_i}{2\rho_i} - \frac{(\nabla\rho_i)^2}{4\rho_i^2} - (\nabla\phi_i)^2\right) - g_{ij}\rho_j + \nu\sum_{j\neq i}\sqrt{\frac{\rho_j}{\rho_i}}\cos(\phi_j-\phi_i) \, .
\end{align}
\end{subequations}
Note in particular that
\begin{equation}
  \hbar\sum_i\dot{\rho}_i = -\hbar^2 \sum_im_i^{-1}\left(\rho_i\nabla^2\phi_i + \nabla\rho_i\cdot\nabla\phi_i\right) = -\hbar^2\sum_i\frac{1}{m_i}\nabla\cdot\left(\rho_i\nabla\phi_i\right) \, ,
\end{equation}
so that in the absence of inhomogeneities, $\sum_i\dot{\rho}_i = 0$ at every point in space.
Accounting for inhomogeneities
\begin{equation}
  \frac{\dd}{\dd t}\int_V\ud{x}{d}\sum_i\rho_i = -\hbar^2\sum_i\frac{1}{m_i}\oint_{\partial V} \rho_i\nabla\phi_i\cdot\dd\vec{S} \, ,
\end{equation}
which follows from the Gross-Pitaevskii equation regardless of the number of condensates.
We recognize
\begin{equation}
  \sum_i \frac{\hbar}{m_i}\rho_i\nabla\phi_i = -\sum_i\frac{i\hbar}{2m_i}\left(\psi_i^*\nabla\psi_i-\psi_i\nabla\psi_i^*\right) \equiv \sum_i{\bf J}_i
\end{equation}
as the total particle current for the condensates.
In the absence of loss of particles through the boundary, the total particle number is conserved.

For simplicity, we redefine the coupling constants
\begin{equation}
  g_{11} = g_{22}  = \gS, \qquad g_{12} = \gC \, .
\end{equation}
As stated in the introduction, in this paper we restrict our considerations to the case of symmetric condensates, with $m_1=m_2\equiv m$ and $g_{11}=g_{22}$.
We will assume these conditions hold in the remainder of the paper,
while the effects of relaxing these assumptions will be explored in an upcoming publication.

Further, it is convenient to work with the following alternative variables:
\begin{equation}
  \label{eqn:phonon-vars}
  \rhoP \equiv \frac{\rho_1 + \rho_2}{2} \qquad \rhoN \equiv \frac{\rho_2-\rho_1}{2} \qquad \phiP \equiv \phi_1+\phi_2 \qquad \phiN \equiv \phi_2-\phi_1 \, ,
\end{equation}
which are (up to a scaling) canonical coordinates.
We can directly express the original variables $\psi_i$ as
\begin{subequations}\label{eqn:psi-phonon-param}
\begin{align}
  \psi_1 &= \sqrt{\rhoP-\rhoN}\, e^{\frac{i}{2}\left(\phiP-\phiN\right)} &
  \psi_2 &= \sqrt{\rhoP+\rhoN}\, e^{\frac{i}{2}\left(\phiP+\phiN\right)} \, .
\end{align}
\end{subequations}
Substituting into the Hamiltonian density and accounting for scaling, we obtain the Hamiltonian density $\mathcal{K}$ for the variables $\rhoP,\rhoN,\phiP,\phiN$:
\begin{align}
  \hbar\mathcal{K} = \mathcal{H} &= \hbar^2\frac{\rhoP}{4m}\left((\nabla\phiP)^2+(\nabla\phiN)^2\right)
  + \hbar^2\frac{\rhoN}{4m}2\nabla\phiP\cdot\nabla\phiN \\
    &+ \frac{\hbar^2}{8m}\left[\left(\rhoP+\rhoN\right)\left(\nabla\ln(\rhoP+\rhoN)\right)^2 + \left(\rhoP-\rhoN\right)\left(\nabla\ln(\rhoP-\rhoN)\right)^2\right] \\
    &+ (\gS+\gC)\rhoP^2 + (\gS-\gC)\rhoN^2 - 2\nu\cos\phiN\sqrt{\rhoP^2-\rhoN^2} \notag \, .
\end{align}
We introduce $\mathcal{K}$ to distinguish it from the function obtained by substituting~\eqref{eqn:psi-phonon-param} directly into~\eqref{eqn:ham-dens-real}.
The overall $\hbar$ factor allows $\mathcal{K}$ be varied directly with respect to~\eqref{eqn:phonon-vars} in Hamilton's equations.

The remainder of our analysis assumes that the coupled GPEs provide an adequate theoretical description of the two-component BEC.
In reality, we expect that at sufficiently short wavelengths additional corrections must be included, with a complete breakdown of the GPE description occurring eventually.
To estimate where this occurs, recall that the four particle contact interactions $\sim g_{ij} \abs{\Psi_i}^2\abs{\Psi_j}^2$ arise as the $S$-wave scattering limit of the full interaction rate, with the potential interactions between the particles (not to be confused with the external trapping potential) approximated as a delta function. Assuming $g_{11}=g_{22}=g$ and $g_{12} = 0$, so that there is only a single scattering length $a_s$, the $S$-wave scattering length $a_s$ is defined by
\begin{equation}
  g = \frac{4\pi\hbar^2 a_s}{m} \, ,
\end{equation}
and the dilute gas approximation requires $\rhoPAve a_s^3 \ll 1$ where $\rhoPAve = \frac{n_1+n_2}{2}$ is the mean particle density.

There are two relevant length scales to consider. The first is the length scale on which fluctuations can probe the internal structure of the interparticle interactions, which occurs when the wavelength of the perturbation is of order the $S$-wave scattering length. This defines a wavenumber,
\begin{equation}
k_{\rm atom} = \frac{2 \pi}{a_s} = \frac{8 \pi^2 \hbar^2}{gm} \, .
\end{equation}
The second important length scale is the ``healing length'', which characterizes the length scale over which the condensate can adjust in the presence of a boundary or defect. The wavenumber $k_{\rm heal}$ associated with the healing length $L_{\rm heal}$ can be estimated by comparing the gradient energy $\sim \hbar^2 k_{\rm heal}^2 / m$ to the interaction term $\sim g \rhoPAve$ in the Hamiltonian, yielding
\begin{equation}
k_{\rm heal} = \sqrt{\frac{g \rhoPAve m}{\hbar^2}} = \sqrt{\frac{\rhoPAve a_s^3}{\pi}} \ k_{\rm atom} \, .
\end{equation}
Note that in the dilute gas approximation, $k_{\rm atom}$ is parametrically larger than $k_{\rm heal}$.

When fluctuations probe the internal structure of the interparticle interactions, we cannot replace these interactions by an effective delta function potential,
and the original Hamiltonian operator~\eqref{eqn:ham} from which the GPE descends is no longer valid.
Therefore, the GPE is not a good description of the system when fluctuations can probe the internal structure of the interparticle interactions.
Meanwhile, the GPE is expected to be a good description on scales below the healing length.  We therefore conclude that the breakdown of the GPE will occur somewhere in the range $k_{\rm heal} < k < k_{\rm atom} $.

\section{Connection to false vacuum decay}\label{sec:effective-action}
We now demonstrate how the cold atom system described above can be connected to a seemingly unrelated physical phenomenon --- relativistic false vacuum decay.
To make this connection, we first demonstrate that under suitable conditions, the phases $\phiP$ and $\phiN$ of the condensates behave as effective relativistic scalar field degrees of freedom.
In the limit $\nuT = 0$, the Lagrangian is invariant under independent global rotations of each phase.
The formation of the condensates $\psi_i$ leads to a spontaneous breakdown of these symmetries, resulting in the presence of two Goldstone modes.
The introduction of $\nuT$ explicitly breaks the $U(1)$ (i.e.\ shift) symmetry associated with the relative phase, generating a sinusoidal potential in the presence of the condensate.
Under a suitable set of experimental conditions (see below), the dynamics of the relative phase variable is described by the sine-Gordon Lagrangian.
At this point, the potential for the relative phase posesses a series of stable degenerate global minima, and unstable degenerate local maxima.
Each of these local extrema is a stationary point in the strictly homogeneous limit, but small perturbations around the local maxima experience tachyonic instabilities, indicating the maxima are not vacua of the system.
The final step to connect to false vacuum decay is to create a series of local potential minima to act as metastable states.
The key insight (c.f.\ Fialko et al~\cite{Fialko:2014xba,Fialko:2016ggg}) is that in the long-wavelength limit the local maxima of the original sinusoidal potential can be converted into local minima through periodic temporal modulation of the coupling $\nuT$.
Explicitly, as we demonstrate below, the relative phase $\phiN$ has a leading order effective Lagrangian of the form
\begin{equation}
  \label{eqn:effective-action-fv}
  \mathcal{L}_{\rm eff}^\phiN \propto \frac{\dot{\phiN}^2}{2} - c_{\rm s}^2\frac{(\nabla\phiN)^2}{2} - V_0\left(-\cos\phiN + \frac{\lPar^2}{2}\sin^2\phiN\right) \, ,
\end{equation}
where $c_{\rm s}$, $V_0$, and $\lPar$ are constants whose explicit form will be given below.
Due to the complexity of intermediate expressions, in the main text we sketch the derivation of the effective action~\eqref{eqn:effective-action-fv}.
We explicitly indicate the key features that lead to a relativistic scalar field description, and the required approximations to obtain the sine-Gordon model.
A detailed derivation is given in~\appref{app:effective-action}.
Finally, we show how modulating $\nuT$ converts the sine-Gordon model into a model suitable for studying false vacuum decay.

\subsection{Effective Relativistic Field Dynamics from Bose Condensates}
It is convenient to apply the path integral formalism for the fluctuation dynamics in the 2-component system.
The astute reader will undoubtedly notice the close parallel between the derivation in this section and the usual transformation from the Hamiltonian to Lagrangian formalism in the path integral language, which illustrates the generality of the construction.
The propagator has the form
\begin{equation}
  \mathcal{Z} = \int \Pi_i\dd\psi_i\dd\psi_i^* e^{\frac{i}{\hbar}\int \dd t\dd^dx \mathcal{L}} \, ,
\end{equation}
where the Lagrangian density is
\begin{equation}
  \label{eqn:lag-full}
  \mathcal{L} = \frac{i\hbar}{2}\left(\psi_i^*\dot{\psi}_i-\psi_i\dot{\psi}_i^*\right) - \mathcal{H} + \chemP\psi_i^*\psi_i \, .
\end{equation}
Here, $\hamD$ is the real Hamiltonian density~\eqref{eqn:ham-dens-real} defined in~\secref{sec:bec-review} and $\chemP$ is the chemical potential, which is a Lagrange multiplier associated with the conservation of total particle number.
Although we work directly in Lorentzian signature, we note that after appropriate Wick rotation we obtain the same results in Euclidean time via the grand partition function.

To obtain an effective action for the phase variables $\phiP$ and $\phiN$, we must integrate out the density variables $\rhoP$ and $\rhoN$.
With this in mind, we expand
\begin{equation}
  \rhoP \equiv \rhoPAve + \pert{\rhoP}({\bf x},t) \equiv \rhoPAve\left(1 + \delRatTot({\bf x},t)\right) \qquad \rhoN({\bf x},t) = \rhoNAve(t) + \pert{\rhoN}({\bf x},t) \, ,
\end{equation}
where we note that the time-independence of $\rhoPAve$ follows from the coupled GPEs.
If we assume the inhomogeneous fluctuations in the local number density are small, we can expand~\eqref{eqn:lag-full} to second order in $\pert{\rhoP}$ and $\pert{\rhoN}$.
We will denote the $i$th order in this expansion by $\lagDdel[i]$.

In order to simplify the result while maintaining the essential aspects of the derivation, we now introduce the additional symmetry assumption $\rhoNAve = 0$.  In the next section we will see this constraint is closely related to the assumption $g_{11}=g_{22}$.
Expanding the Lagrangian, we obtain
\begin{subequations}
\begin{align}
  \lagDdel[0] &= -\hbar\rhoPAve\dot{\phiP} - \frac{\hbar^2\rhoPAve}{4m}(\nabla\phiP)^2 - \frac{\hbar^2\rhoPAve}{4m}(\nabla\phiN)^2 + 2\nuT\rhoPAve\cos\phiN - (\gS+\gC)\rhoPAve^2 + 2\chemP \rhoPAve \\
  \lagDdel[1] &= \left(\begin{array}{c}
                -\hbar\dot{\phiP} - 2(\gS+\gC)\rhoPAve + 2\nu\cos\phiN - \frac{\hbar^2}{4m}(\nabla\phiP)^2 - \frac{\hbar^2}{4m}(\nabla\phiN)^2 + 2\chemP \\
                -\hbar\dot{\phiN} - \frac{\hbar^2}{2m}\nabla\phiP\cdot\nabla\phiN
                 \end{array}\right)^T
           \left(\begin{array}{c} \pert{\rhoP} \\ \pert{\rhoN} \end{array}\right) \\
  \lagDdel[2] &= -\frac{1}{2}\left(\begin{array}{c} \pert{\rhoP} \\ \pert{\rhoN} \end{array}\right)^T
             \left(\begin{array}{cc} 2(\gS+\gC) +  \frac{\hbar^2\nabla^T\cdot\nabla}{2\rhoPAve m} & 0 \\
                       0 & 2(\gS-\gC) + 2\frac{\nu}{\rhoPAve}\cos\phiN + \frac{\hbar^2\nabla^T\cdot\nabla}{2\rhoPAve m} \end{array}\right)
             \left(\begin{array}{c} \pert{\rhoP} \\ \pert{\rhoN} \end{array}\right) \, .
\end{align}
\end{subequations}
Since the Lagrangian density is quadratic in $\pert{\rhoP}$ and $\pert{\rhoN}$, we can perform the Gaussian integrals to obtain an effective Lagrangian for the phase variables $\phiP$ and $\phiN$.
From the viewpoint of obtaining a relativistic scalar field theory, the key observation is that the $\hbar\dot{\phiP}\pert{\rhoP}$ and $\hbar\dot{\phiN}\pert{\rhoN}$ terms lead to contributions $\sim\dot{\phiP}^2$ and $\sim\dot{\phiN}^2$ in this effective Lagrangian density.
Combining this with the gradient and potential energy contributions already present in $\lagDdel[0]$, we see that modulo various environmental corrections and distortions to the dispersion relationship, we will obtain a Lagrangian for a relativistic scalar field.
To obtain the sine-Gordon model for $\phiN$, we now make the following set of approximations (details can be found in~\appref{app:effective-action}):
\begin{enumerate}
  \item we keep only the leading order gradient energies in $\left(\nabla\phiN\right)^2$, $\left(\nabla\phiP\right)^2$, and $\nabla\phiN\cdot\nabla\phiP$,
  \item we assume the experimental parameters satisfy $\nuT/(\gS\pm\gC)\rhoPAve \ll 1$,
  \item we work in the long-wavelength limit and take $\hbar^2k^2/(\gC\pm\gS)m\rhoPAve \ll 1$ (e.g. on scales larger than the healing length),
  \item we ignore backreaction of the fluctuations onto the homogeneous background, and
  \item we ignore corrections induced by interactions between $\phiN$ and $\phiP$ by truncating rather than integrating out the dynamics of $\phiP$.
\end{enumerate}
Combined with the assumptions of a homogeneous background, $\rhoNAve=0$, and smallness of $\pert{\rhoP}$ and $\pert{\rhoN}$, we obtain the following effective Lagrangian for $\phiN$:
\begin{equation}
  \label{eqn:lag-eff-sine-gordon}
  \frac{2(\gS-\gC)}{\hbar^2}\mathcal{L}_{\rm eff}^\phiN \approx \frac{\dot{\phiN}^2}{2} - \frac{(\gS-\gC)\rhoPAve}{m}\frac{(\nabla\phiN)^2}{2} + \frac{4\nuT\rhoPAve(\gS-\gC)}{\hbar^2}\cos\phiN \, ,
\end{equation}
which is of course the Lagrangian for the sine-Gordon model.
In order to determine the precise relationship between the analog false vacuum dynamics and the BEC experiment, one should carefully account for the various correction terms as well as integrate out the dynamics of $\phiP$.
We leave a detailed investigation of the effects of the environmental corrections and comparison to numerical simulations to future work.

Given the many approximations made to establish this analogy, we will be able to use the 2-component BEC not only to test the universality of the false vacuum decay, but also its robustness. At the same time one needs to be careful when extracting quantitative information, e.g.~decay rates, from this system.
In particular, even if the approximations hold for short time-scales, errors may accumulate over time if the decay rates are too small.
Detailed analysis is required to compare the time-scale for this accumulation to the typical lifetime of the condensate, which is beyond the scope of this paper.

\subsection{Generation of False Vacuum Minima Through Modulated Coupling}
From the above analysis, we see that the introduction of the particle conversion parameterized by $\nuT$ induces an effective sinusoidal potential into the dynamics of the relative phase degree of freedom.
Up to higher derivative corrections (not included above), non-canonical contributions to the kinetic term (which are small in the long-wavelength regime and for $\nuT \ll (\gS-\gC)\rhoPAve$), and some (possibly nonlocal) corrections from interaction with the environment, we showed that the relative phase variable $\phiN$ has the effective Lagrangian of the sine-Gordon model.
Following~\cite{Fialko:2014xba,Fialko:2016ggg}, we now demonstrate that the local maxima of cosine potential can be transformed into local minima for the long-wavelength degrees of freedom through an appropriate modulation of the coupling $\nuT$:
\begin{equation}
  \label{eqn:nu-t-dep}
  \nuT(t) = \nuZero + \nuAmp\hbar\nuFreq\cos(\nuFreq t) \, .
\end{equation}
Rather than integrate out this time-dependence directly in the propagator, we will first derive an effective time-averaged Hamiltonian for the full system, and use this to derive modifications to the quadratic Lagrangian presented above~\cite{PhysRevX.4.031027}.
Given a Hamiltonian density of the form
\begin{equation}
  \mathcal{H} = \mathcal{H}_0 + \Delta\mathcal{H}\cos(\nuFreq t) \, ,
\end{equation}
in the absence of instabilities in the short-wavelength modes that can destabilize the system,
modes with characteristic time-scales much less than $\omega^{-1}$ will evolve in the effective Hamiltonian density
\begin{equation}
  \mathcal{H}_{\rm eff} = \mathcal{H}_0 + \frac{1}{4(\hbar\nuFreq)^2}[[\Delta\mathcal{H},\mathcal{H}_0],\Delta\mathcal{H}] + \mathcal{O}((\hbar\nuFreq)^{-4}) \equiv \mathcal{H}_0 + \Delta\mathcal{H}_{\rm eff} \, .
\end{equation}
At each subsequent order of the expansion in $\omega^{-1}$, an additional commutator with $\Delta\mathcal{H}$ is introduced.  Heuristically, the new factor of $\Delta\mathcal{H}$ introduces a factor of $\rhoPAve\hbar\omega$, while the action of the commutator differentiates each sinusoidal function and reduces the power of $\rhoPAve$ by one.  This means that for the parameterization~\eqref{eqn:nu-t-dep} the expansion in $\omega^{-1}$ becomes an expansion in $\nuAmp$.
For the purposes of illustration, here we ignore issues of operator ordering and use the correspondence between quantum commutators and classical Poisson brackets
\begin{equation}
  [[\Delta\op{\mathcal{H}},\op{\mathcal{H}}_0],\Delta\op{\mathcal{H}}] \to (i\hbar)^2\{\{\Delta\mathcal{H},\mathcal{H}_0\},\Delta\mathcal{H}\}_{\rm PB} + \mathcal{O}(\hbar^3) \, ,
\end{equation}
where $\op{\mathcal{H}}$ indicates the quantum operator and $\mathcal{H}$ the corresponding classical function.
For the case considered here, we have
\begin{equation}
  \hbar\Delta\mathcal{H} = -2\nuAmp\hbar\nuFreq\cos\phiN\sqrt{\rhoP^2-\rhoN^2} 
\end{equation}
where we have included the $\hbar$ to account for the fact that our preferred variables are scaled canonical coordinates.
In the Poisson bracket approximation, the contribution from the gradient terms vanish.
Therefore, we consider only the potential contributions.
A straightforward calculation gives
\begin{align}
  \label{eqn:ham-tave-correction}
  \nuAmp^{-2}\Delta\mathcal{H}_{\rm eff} &\approx -2(\gS-\gC)\rhoN^2 + 2(\gS-\gC)\left(\rhoP^2-\rhoN^2\right)\sin^2\phiN \, ,
\end{align}
which can be easily (but tediously) verified by working directly with the variables $\psi_i$ and $\psi_i^*$ instead.

\begin{figure}
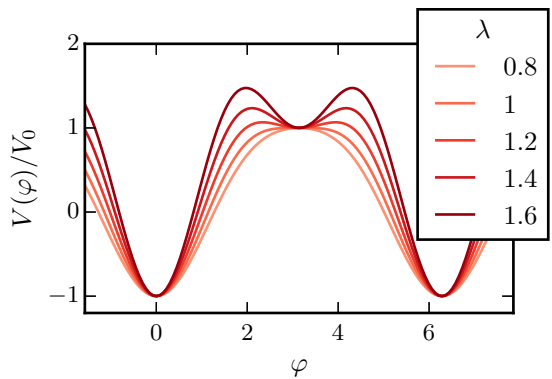

  \centering
  \includegraphics[width=2.97in]{{{effective-potentials}}}
  \caption{The tree-level time-averaged effective potential~\eqref{eqn:pot-eff-tree} for the slow-time dynamics of the effective scalar field $\phiN$.  For simplicity, we have assumed equal condensate densities, so that $\rhoNAve = 0$.}
  \label{fig:effective-potential-tree}
\end{figure}

We now consider how this correction to $\mathcal{H}$ modifies the effective action presented above.
As before, we assume $\rhoNAve = 0$ and $m_1=m_2$.
Further, in order to most directly relate to the previous analysis, we make the same sequence of approximations as above.
The time-averaged analog of~\eqref{eqn:lag-eff-sine-gordon} is
\begin{align}\label{eqn:lag-eff-tave}
  \frac{2(\gS-\gC)(1-2\nuAmp^2)}{\hbar^2}\bar{\mathcal{L}}_{\rm eff}^{\phiN} &= \frac{1}{2}\dot{\phiN}^2 - \frac{(\gS-\gC)(1-2\nuAmp^2)\rhoPAve}{m}\frac{1}{2}(\nabla\phiN)^2 \notag \\
  &- \frac{4\nuZero\rhoPAve(\gS-\gC)(1-2\nuAmp^2)}{\hbar^2}\left(-\cos\phiN + \frac{\nuAmp^2(\gS-\gC)\rhoPAve}{\nuZero}\sin^2\phiN\right) \, ,
\end{align}
where we have used $\bar{\mathcal{L}}$ to indicate the effects of the time-averaging.
We thus see that the introduction of high frequency oscillations in $\nuT$ has led to the creation of a stabilizing effective potential for the slow dynamics of the system.
Comparing with~\eqref{eqn:effective-action-fv} we identify the sound speed
\begin{equation}
  \label{eqn:c-sound-tave}
  c_{\rm s}^2 = \frac{(\gS-\gC)\rhoPAve(1-2\nuAmp^2)}{m}\, .
\end{equation}
Note that our expansion is only valid for $\nuAmp \ll 1$, so that~\eqref{eqn:c-sound-tave} does not imply the oscillations can induce an imaginary sound speed. 
At tree level, the potential has the form
\begin{equation}
  \bar{V}_{\rm tree} =  V_0\left(-\cos\phiN + \frac{\lPar^2}{2}\sin^2\phiN\right) \, ,
  \label{eqn:pot-eff-tree}
\end{equation}
where we have defined
\begin{equation}
  \label{eqn:pot-params}
  V_0 = \frac{4\nuZero\rhoPAve(\gS-\gC)(1-2\nuAmp^2)}{\hbar^2} \qquad {\rm and} \qquad \lPar^2 = \frac{2\nuAmp^2(\gS-\gC)\rhoPAve}{\nuZero} \, .
\end{equation}
The effective potential~\eqref{eqn:pot-eff-tree} is illustrated in~\figref{fig:effective-potential-tree}.
In the remainder of the paper we will explore the validity of this approximation, specifically the assumption that the high temporal frequency dynamics is stable and can be suitably averaged to obtain the effective action~\eqref{eqn:lag-eff-tave}.

Unfortunately, some of the elegance of the above derivation is spoiled when we include the higher derivative corrections.
Further complications arise if we also break the symmetry between the two condensates, by allowing either $\rhoNAve \neq 0$ or $m_1\neq m_2$.
We leave exploration of the precise connection between relativistic field theory and symmetry properties of the BECs to future work.

\subsection{Effective Action for Linear Phase Perturbations}
The effective Lagrangians~\eqref{eqn:lag-eff-sine-gordon} and~\eqref{eqn:lag-eff-tave} are valid for the \emph{nonlinear} dynamics of $\phiN$ subject to the assumptions outlined in~\appref{app:effective-action} and summarized above.
Due to its nonlinear nature, we can use~\eqref{eqn:lag-eff-tave} to infer the existence of a metastable false vacuum in the effective relativistic dynamics of $\phiN$.

However, to connect with the linear fluctuation analysis presented in the remainder of the paper, it is useful to instead obtain the effective Lagrangians for linear fluctuations in $\phiN$ and $\phiP$.
We therefore decompose
\begin{equation}
  \phiP({\bf x},t) = \phiPAve(t) + \pert{\phiP}({\bf x},t) \qquad \phiN(({\bf x},t) = \phiNAve(t) + \pert{\phiN}({\bf x},t)
\end{equation}
and work to quadratic order in the full set of fluctuation variables.
As will be seen in~\secref{sec:homogeneous-bg}, we are interested in states with ${\cos\phiNAve = \pm 1 \equiv \cphi}$, so we impose this additional constraint.
In contrast to~\eqref{eqn:lag-eff-sine-gordon}, we make no assumptions about the magnitude of ${\nu/(\gS-\gC)\rhoPAve}$ or ${\hbar^2k^2/4m\rhoPAve(\gS\pm\gC)}$.
The details of this expansion are provided in~\appref{app:effective-action}, where we see that the total and relative phase fluctuations decouple provided $\rhoNAve = 0$, $g_{11}=g_{22}$ and $m_1=m_2$.
In particular, we obtain the following effective Lagrangian for $\pert{\phiN}$
\begin{align}
  \label{eqn:lag-linear}
  \frac{2(\gS-\gC)}{\hbar^2}\mathcal{L}_{\rm quad}^\phiN &= \frac{1}{2}\pert{\dot{\phiN}}\left(1 + \frac{\cphi\nu}{(\gS-\gC)\rhoPAve} - \frac{\hbar^2}{4(\gS-\gC)\rhoPAve m}\nabla^2\right)^{-1}\pert{\dot{\phiN}} \notag \\
  &- \frac{(\gS-\gC)\rhoPAve}{m}\frac{(\nabla\pert{\phiN})^2}{2} - \frac{4\cphi\nu\rhoPAve(\gS-\gC)}{\hbar^2}\frac{\pert{\phiN}^2}{2} \, .
\end{align}
As we will see below, the equation of motion derived from this effective Lagrangian matches that obtained from direct manipulation of the equations of motion~\eqref{eqn:wave-eqn-rel-phase-sym}, indicating that our effective action approach reproduces the correct behavior in the regime of linear fluctuations.

We can similarly incorporate~\eqref{eqn:ham-tave-correction} to obtain an effective Lagrangian for linear perturbations of the relative phase evolving with the effective time-averaged Hamiltonian
\begin{align}
  \label{eqn:lag-teff-linear}
  \frac{2(\gS-\gC)}{\hbar^2}\bar{\mathcal{L}}_{\rm quad}^\phiN &= \frac{1}{2}\pert{\dot{\phiN}}\left(1-2\nuAmp^2 + \frac{\cphi\nuZero}{(\gS-\gC)\rhoPAve} - \frac{\hbar^2}{4(\gS-\gC)\rhoPAve m}\nabla^2\right)^{-1}\pert{\dot{\phiN}} \notag \\
  &- \frac{(\gS-\gC)\rhoPAve}{m}\frac{(\nabla\pert{\phiN})^2}{2} - \frac{4\nuZero\rhoPAve(\gS-\gC)}{\hbar^2}\left(\cphi+\lPar^2\right)\frac{\pert{\phiN}^2}{2} \, ,
\end{align}
which we will use to make analytic predictions for the growth rates of the homogeneous mode below.

\section{Homogeneous Background Solutions}\label{sec:homogeneous-bg}
Before proceeding with the analysis of fluctuations, we must first find a background to expand around.
In order to find suitable states, we search for eigenstates of the (nonlinear) Hamiltonian
\begin{equation}
  \hat{H}[\vec{\wf}] = \mu\vec{\wf} \implies i\hbar\dot{\wf}_i = \mu\wf_i \, ,
\end{equation}
where $\vec{\wf} \equiv \left(\begin{array}{cc} \wf_1 & \wf_2 \end{array}\right)^T$.
We can absorb the constant $\mu$ into a global phase $\tilde{\wf}_i = \wf_ie^{i\mu t}$ with $\dot{\tilde{\wf}}_i = 0$.
Alternatively, we can evolve the condensates with a modified Hamiltonian
\begin{equation}
  F = H - \chemPBG \int\ud{x}{d}\sum_i\abs{\wf_i}^2 = H - 2\chemPBG \int\ud{x}{d}\rhoP
\end{equation}
and search for zero eigenmodes of $F$.

For simplicity, we assume a homogeneous condensate, which is consistent with the equations of motion in the absence of a trapping potential in the GPE as we have assumed here.
It is convenient to define an effective potential
\begin{equation}
  \potBG \equiv \frac{F}{V} \, ,
\end{equation}
where $V$ is the spatial volume of the system.

The variables~\eqref{eqn:phonon-vars} introduced in~\secref{sec:bec-review} provide a convenient basis to study the homogeneous dynamics.
The Hamiltonian $K$ for these variables is
\begin{equation}
  \label{eqn:eff-pot-bg}
  \hbar K = \potBG = \left(\gS+\gC\right)\rhoBG^2 + (\gS-\gC)\dRhoBG^2 - 2\nu\cos\dphiBG\sqrt{\rhoBG^2 - \dRhoBG^2} - 2\chemP\rhoBG \, .
\end{equation}
Here, $\tphiBG$ and $\rhoBG$ form a canonically conjugate pair, as do $\dRhoBG$ and $\dphiBG$,
and the stability properties are easily analyzed from the Hessian of~\eqref{eqn:eff-pot-bg}.

For notational simplicity, it is convenient to define
\begin{align}
  \nuBG &\equiv \frac{\nuT}{\rhoBG(\gS-\gC)} \, ,
\end{align}
as well as the dimensionless dynamical parameter
\begin{equation}
  \drhoRat \equiv \frac{\dRhoBG}{\rhoBG} \, .
\end{equation}

To simplify some notation, we now arrange our variables into the vector
\begin{equation}
  \label{eqn:bg-var-vector}
  {\bf q} = \left[\begin{array}{cccc} \rhoBG & \dRhoBG & \tphiBG & \dphiBG \end{array}\right]^T \, .
\end{equation}
It is easily verified that the equations of motion can be obtained by extremizing $\potBG$,
so that stationary solutions occur when $\nabla_{\bf q}\potBG = 0$, where the derivative is with respect to our variables~\eqref{eqn:bg-var-vector}.
Stationarity of $\dRhoBG$ follows from extremization with respect to $\dphiBG$,
\begin{equation}
  \label{eqn:dVdPhi}
  \hbar\dot{\dRhoBG} = \frac{\partial\potBG}{\partial\dphiBG} = 0 = 2\nu\sin\dphiBG\sqrt{\rhoBG^2+\dRhoBG^2} \, .
\end{equation}
Concentrating on the states where both condensates are occupied, we thus see that at the stationary points
\begin{equation}
  \cos\dphiBG = \pm 1 \equiv \cphi \, .
\end{equation}
Next, enforcing stationarity in $\dphiBG$ gives us an equation for the relative density difference $\dRhoBG$:
\begin{align}
  \label{eqn:dVdDelRho}
  \hbar\dot{\dphiBG} = -\frac{\partial\potBG}{\partial\dRhoBG} = 0 &= -2\dRhoBG(\gS-\gC)\left(1 + \frac{\nu}{\rhoBG(\gS-\gC)}\cos\dphiBG\left(1-\left(\frac{\dRhoBG}{\rhoBG}\right)^2\right)^{-1/2}\right) \, .
\end{align}
Finally, the chemical potential $\chemP$ follows from the stationarity of $\tphiBG$:
\begin{align}
  \hbar\dot{\tphiBG} = -\frac{\partial\potBG}{\partial\rhoBG} &= 0 = -2(\gS+\gC)\rhoBG + 2\nu\cos\dphiBG\left(1-\left(\frac{\dRhoBG}{\rhoBG}\right)^2\right)^{-1/2} + 2\chemP \, .
\end{align}

To determine the stability of these solutions, consider small deviations $\pert{{\bf q}}$ around a point ${\bf q}_0$ which satisfy
\begin{equation}
  \frac{\dd\pert{q}_i}{\dd t} = D_{ij}[{\bf q}_0]\pert{q}_j \, ,
\end{equation}
with the matrix
\begin{equation}
  D_{ij} = J_{ik}\frac{\partial^2\potBG}{\partial q_k\partial q_j}
\qquad \mathrm{and} \qquad
  J = \left[\begin{array}{cc} 0 & \mathbb{I}_2 \\ -\mathbb{I}_2 & 0 \end{array}\right] \, .
\end{equation}
Since $\partial_{\tphiBG} \potBG = 0$, we can consider $(\dRhoBG,\dphiBG)$ in isolation, as explicitly seen from the characteristic equation for $D_{ij}$:
\begin{equation}
  \eig^2\left(\eig^2 + (\partial_{\dphiBG\dphiBG}\potBG\partial_{\dRhoBG\dRhoBG}\potBG - \partial_{\dphiBG\dRhoBG}^2\potBG)\right) = 0 \, .
\end{equation}
Unfortunately, due to the degeneracy of the Hamiltonian, the simplistic approach given here cannot address the full stability of the system (since the $\eig = 0$ eigenvalues are degenerate).  However, these eigenvalues are associated with the $(\rhoBG,\tphiBG)$ dynamics, which are seen to possess no dynamical instabilities in the homogeneous limit.

The required components of the Hessian are:
\begin{subequations}
\begin{align}
  \frac{\partial^2\potBG}{\partial\dphiBG^2} &= 2\rhoBG\nu\cos\dphiBG\sqrt{1-x^2}\, , \\
  \frac{\partial^2\potBG}{\partial\dRhoBG^2} &= 2(\gS-\gC)\left(1 + \frac{\nu}{\rhoBG(\gS-\gC)}\cos\dphiBG\left(1 - x^2\right)^{-3/2}\right)\, , \\
  \frac{\partial^2\potBG}{\partial\dphiBG\partial\dRhoBG} &= -2\rhoBG\frac{\nu}{\rhoBG}\sin\dphiBG\left(1-x^2\right)^{-1/2} \, .
\end{align}
\end{subequations}
Thus around our states of interest with $\sin\dphiBG=0$, we have $\partial_{\dphiBG\dRhoBG}\potBG = 0$.
As outlined in~\secref{sec:effective-action}, we are interested in cases when the false vacuum minima for $\dphiBG$ arise from the local potential maxima after the introduction of temporal modulations in $\nu$.
We are therefore interested in unstable states (i.e.\ $\eig^2 > 0$) with $\partial_{\dphiBG\dphiBG}\potBG < 0$.
The latter condition immediately gives
\begin{equation}
  \label{eqn:phi-unstable}
  \nuBG\sigma < 0 \, ,
\end{equation}
while the former tells us
\begin{equation}
  \label{eqn:fv-unstable}
  \frac{\partial^2\potBG}{\partial\dRhoBG^2} > 0 \, .
\end{equation}

In the $\dRhoBG = 0$ state, the density and chemical potential are then related by
\begin{equation}
  \chemP = \rhoBG(\gS+\gC) - \cphi\nuT \, ,
\end{equation}
with $\cphi = \cos\dphiBG = \pm 1$.  $\cphi\nuT > 0$ corresponds to the true vacuum, and $\cphi\nuT < 0$ to the false vacuum.

\section{Linear Fluctuation Dynamics about the False Vacuum}\label{sec:linear-symmetric}
Having obtained the relevant background solutions for the condensates, we now turn to the question of their stability for non-zero wavenumbers.
Since we are interested in the possible destabilization of short wavelength modes from the oscillations in $\nuT$, we will work directly with the full-time dependent equations of motion, rather than those derived from the time-averaged Hamiltonian~\eqref{eqn:lag-teff-linear}.
In~\appref{app:linear} we derive the equations of motion for linear perturbations about an arbitrary homogeneous background in a symmetric two condensate experiment.

Here we want to study fluctuations around the proposed false vacuum.
From~\secref{sec:homogeneous-bg} this implies $\dRhoBG = 0$, $\sin\phiNAve = 0$, and $\nu\cos\phiNAve < 0$.
Imposing the former two background conditions in~\eqref{eqn:lin-eqns-bg-tdep}, we see that the perturbation variables $\pert{\rhoP}$ and $\pert{\phiP}$ decouple from $\pert{\rhoN}$, and $\pert{\phiN}$.
It is also convenient to work with the dimensionless density perturbations
\begin{equation}
  \delRatTot \equiv \frac{\pert{\rhoP}}{\rhoPAve} \qquad \delRatDiff \equiv \frac{\pert{\rhoN}}{\rhoPAve} \, .
\end{equation}

The equations for the total density ($\delRatTot$) and phase ($\pert{\phiP}$) perturbations are
\begin{subequations}\label{eqn:phonon-total-symmetric}
\begin{align}
  \hbar\frac{\dd\delRatTot}{\dd t} &= -\frac{\hbar^2}{2m}\nabla^2\pert{\phiP} \\
  \hbar\frac{\dd\pert{\phiP}}{\dd t} &= \frac{\hbar^2}{2m}\nabla^2\delRatTot - 2(\gS+\gC)\rhoPAve\delRatTot \, .
\end{align}
\end{subequations}
From this we immediately read off the dispersion relationship
\begin{equation}
  \label{eqn:dispersion-tot-phase}
  \hbar^2\omega_{k,{\rm tot}}^2 = \frac{\hbar^2k^2}{2m}\left(\frac{\hbar^2k^2}{2m}+2\rhoPAve(\gS+\gC)\right) \, ,
\end{equation}
and the sound speed for the low-$k$ modes
\begin{equation}\label{cs2-total}
  \csTot^2 = \frac{\rhoPAve(\gS+\gC)}{m} = \frac{\gS\rhoPAve}{m}\left(1+\frac{\gC}{\gS}\right) \, .
\end{equation}
Using these, we can rewrite~\eqref{eqn:phonon-total-symmetric} as a wave equation for $\delPhTot$
\begin{equation}
  \label{eqn:wave-eqn-tot-phase-sym}
  \frac{\partial^2\delPhTot}{\partial t^2} - \csTot^2\nabla^2\delPhTot + \left(\frac{\hbar}{(\gS+\gC)\rhoPAve}\right)^2\frac{\csTot^4}{4}\nabla^4\delPhTot = 0 \, .
\end{equation}
This of course matches the equation of motion derived from our second order effective Lagrangian~\eqref{eqn:eom-tot-phase-lag}.
For $\gS+\gC < 0$ (i.e.\ an imaginary sound speed), there is a tachyonic instability for the modes with $\hbar^2k^2 \leq 4m\rhoPAve(\gS+\gC)$.

The situation is more interesting for the relative density ($\delRatDiff$) and phase ($\pert{\phiN}$) phonons,
as they couple to each other through $\nuT$:
\begin{subequations}
  \label{eqn:phonon-rel-symmetric}
\begin{align}
  \hbar\frac{\dd\delRatDiff}{\dd t} &= -\frac{\hbar^2}{2m}\nabla^2\pert{\phiN} + 2\cphi\nuT\pert{\phiN} \\
  \hbar\frac{\dd\pert{\phiN}}{\dd t} &= \frac{\hbar^2}{2m}\nabla^2\delRatDiff - 2\rhoPAve(\gS-\gC)\delRatDiff - 2\cphi\nuT\delRatDiff \, .
\end{align}
\end{subequations}
In this case, the dispersion relationship is
\begin{equation}
  \label{eqn:dispersion-rel-phase-symmetric}
  \hbar^2\omega_{k,{\rm rel}}^2 = \left(\frac{\hbar^2k^2}{2m} + 2\nuT\cphi\right)\left(\frac{\hbar^2k^2}{2m} +2\rhoPAve(\gS-\gC)+2\nuT\cphi \right) \, .
\end{equation}
For $\nuT = 0$, the low-$k$ modes thus propagate with sound speed
\begin{equation}
  \label{eqn:cs2-relative}
  \csRel^2 = \frac{\rhoPAve(\gS-\gC)}{m} = \frac{\rhoPAve\gS}{m}\left(1-\frac{\gC}{\gS}\right)  \, .
\end{equation}
In the absence of $\nuT$, we see that the effect of a non-zero cross-coupling $\gC$ is to split the total and relative perturbations into two distinct groups characterized by their sound speeds.
On the other hand, for $\gS-\gC > 0$ the $\cphi\nuT > 0$ states become gapped, with an effective zero-mode frequency
\begin{equation}
    \hbar^2\Omega^2_{\rm eff} = 4\cphi\nuT(\gS-\gC)\rhoPAve\left(1+\frac{\cphi\nuT}{\rhoPAve(\gS-\gC)} \right) \, ,
\end{equation}
while the $\cphi\nuT < 0$ states instead develop a tachyonic instability for $\hbar^2k^2 < 4m\abs{\nuT}$.

Finally, if $\gS-\gC < 0$, an additional tachyonic instability appears for modes with $\hbar^2k^2 < 4m\rhoPAve\abs{\gS-\gC} - 4m\cphi\nuT$ in the $\cphi\nuT > 0$ phases.
However, in the $\cphi\nuT < 0 $ phases, the modes with $\hbar^2k^2 < -4m\cphi\nuT$ are stabilised, while the modes with $-4m\cphi\nuT < \hbar^2k^2 < -4m\cphi\nuT + 4mn\abs{\gS-\gC}$ develop a tachyonic instability.

In order to most directly connect with the analog scalar field dynamics, it is convenient to convert these coupled first order equations into a second order equation for $\pert{\phiN}$.
Allowing $\nu = \nu(t)$ while maintaining $\cos\phiNAve=\cphi=\pm 1$, we have
\begin{align}
  \label{eqn:wave-eqn-rel-phase-sym}
    \frac{\partial^2\delPhDiff}{\partial t^2} &- \frac{\rhoPAve(\gS-\gC)}{m}\left(1+2\frac{\cphi\nuT}{\rhoPAve(\gS-\gC)}\right)\nabla^2\delPhDiff \notag \\
  &+ \left(\frac{2n(g-g_c)}{\hbar}\right)^2\frac{\cphi\nuT}{\rhoPAve(\gS-\gC)}\left(1+\frac{\cphi\nuT}{\rhoPAve(\gS-\gC)}\right)\delPhDiff  \\
    &-\frac{\cphi}{(\gS-\gC)\rhoPAve}\frac{\dd\nuT}{\dd t}\left(1 - \frac{\hbar^2}{4m\rhoPAve(\gS-\gC)}\nabla^2 + \frac{\cphi\nuT}{\rhoPAve(\gS-\gC)}\right)^{-1}\frac{\partial\dphaseD}{\partial t}  +\frac{\hbar^2}{4m^2}\nabla^4\dphaseD = 0 \notag \, . 
\end{align}
This is equivalent to the Euler-Lagrange equations~\eqref{eqn:eom-rel-phase-lag} for the effective Lagrangian~\eqref{eqn:lag-eff-linear-rel} obtained from working to quadratic order in all of our perturbation variables.

At linear order, the dynamics of $\delPhDiff$ thus differ from that of a free massive relativistic scalar field in Minkowski space through: the presence of a friction term $\partial_t\delPhDiff$, a modified dispersion induced by the $\nabla^4$ term, and the time-dependence of the Laplacian term through $\nuT(t)$.
Furthermore, owing to the presence of the inverse Laplacian, the equations written in second order form are in general nonlocal and depend on the boundary conditions.
All three of these corrections can ultimately be traced back to the non-canonical multiplier of the field kinetic energy term in~\eqref{eqn:lag-linear}.
In taking the standard sine-Gordon limit, we assumed that the gradient energy of the density fluctuations could be neglected, and that $\nuZero/\rhoPAve(\gS-\gC) \ll 1$.  The effect of these two approximations is to discard the non-canonical multiplier, so that these corrections do not appear in the equations derived from~\eqref{eqn:lag-eff-sine-gordon}.
We expect the linear dynamics to be different from a relativistic field in Minkowski space because of the explicit time-dependence of the system at hand.
In fact, in the weak resonance regime and in the absence of a potential, we can engineer an effective field in an FRW spacetime, see for example~\cite{Jain:2007gg,Weinfurtner:2008if}.

\subsection{Floquet Formalism}
In the previous subsection we looked at the linear stability of perturbations around homogeneous states with $\phiNAve = i\pi$ ($i \in \mathbb{Z}$) that apply when $\nuT$ is time independent, in which case it is sufficient to consider the dispersion relationship.
For $\nuT = 0$, we saw that no instabilities were present provided $\gS\pm\gC > 0$.
On the other hand, for $\cphi\nuT < 0$ a tachyonic instability is induced in the coupled fluctuations between the relative density and relative phase perturbations.
Now we analyze the effects of introducing a periodic modulation
\begin{equation}
  \nuT = \nuT(t) = \nuZero + \nuAmp\hbar\nuFreq\cos(\nuFreq t) \, .
\end{equation}
As seen above, around the stationary background solutions, $\pert{\rhoP}$ and $\pert{\phiP}$ are decoupled from the dynamics of $\pert{\rhoN}$ and $\pert{\phiN}$.
Furthermore, only $\pert{\rhoN}$ and $\pert{\phiN}$ feel the presence of $\nuT$.
It is therefore sufficient to consider the perturbations $\pert{\rhoN}$ and $\pert{\phiN}$ in isolation.
For convenience, we introduce the dimensionless parameters
\begin{equation}
  \label{eqn:dimensionless-vars}
  \nuBarSym \equiv \frac{\nuZero}{\rhoPAve(\gS-\gC)} \qquad \wBarSym \equiv \frac{\hbar\omega}{\rhoPAve(\gS-\gC)} \qquad \kBarSym \equiv \sqrt{\frac{\rhoPAve(\gS-\gC)}{m}}\frac{\hbar k}{\rhoPAve(\gS-\gC)} \, .
\end{equation}

Assuming the experiment can be treated as spatially periodic, we obtain the following equations for the Fourier modes ($\pert{\tilde{\zeta}}_{\bf k}$ and $\pert{\tilde{\phiN}_{\bf k}}$) in terms of the time coordinate $\tau = \nuFreq t$:
\begin{equation}\label{eqn:linear-fourier-modes}
  \frac{\dd}{\dd\tau}\left(\begin{array}{c} \drhoDK \\ \dphaseDK \end{array}\right) =
    \frac{2}{\wBarSym}\left[\begin{array}{cc} 0 & \left(\frac{\kBarSym^2}{4} + \cphi\nuBarSym\right) + \nuAmp\wBarSym\cphi\cos\tau \\
                            -\left(\frac{\kBarSym^2}{4} + \cphi\nuBarSym\right) - 1 - \nuAmp\wBarSym\cphi\cos\tau & 0 \end{array}\right]
                     \left(\begin{array}{c} \drhoDK \\ \dphaseDK \end{array}\right) \, .
\end{equation}

We perform our analysis using~\eqref{eqn:linear-fourier-modes}, but to connect with the usual discussions of parametric resonance in field theory, we note that this can be rewritten as the following damped wave equation for the phase alone:
\begin{equation}
  \ddot{\tilde{\dphaseD}}_{\bf k} +  \Gamma_k(\tau)\dot{\tilde{\dphaseD}}_{\bf k} + M_k^2(\tau)\dphaseDK = 0 \, ,
\end{equation}
with
\begin{subequations}
\begin{align}
  \Gamma_k(\tau) &= \cphi\wBarSym\nuAmp\left(\frac{\kBarSym^2}{4} + \cphi\nuBarSym + 1 + \nuAmp\wBarSym\cphi\cos\tau\right)^{-1}\sin\tau \notag \\
      &= -\frac{\dd}{\dd\tau}\ln\left(\frac{\kBarSym^2}{4}+\cphi\nuBarSym + 1 + \nuAmp\wBarSym\cphi\cos\tau\right)  \\
  M_{k}^2(\tau) &= \frac{4}{\wBarSym^2}\left(\frac{\kBarSym^2}{4}+\cphi\nuBarSym+1\right)\left(\frac{\kBarSym^2}{4}+\cphi\nuBarSym\right) + 2\nuAmp^2 \notag \\
  &+ \frac{2\nuAmp}{\wBarSym}\left[4\cphi\nuBarSym + 2 + \kBarSym^2\right]\cphi\cos\tau
  + 2\nuAmp^2\cos2\tau \, .
\end{align}
\end{subequations}
Alternatively, we can define
\begin{equation}
  \pert{\bar{\phiN}}_{\bf k} \equiv  \left(\frac{\kBarSym^2}{4}+\cphi\nuBarSym+1+\nuAmp\wBarSym\cphi\cos\tau\right)^{-1/2}\dphaseDK\, ,
\end{equation}
which satisfies
\begin{equation}
  \pert\ddot{\bar{\phiN}} + \bar{M}_k^2\pert{\bar{\phiN}} = 0
\end{equation}
with
\begin{align}
  \bar{M}_k^2 = M_k^2 - \frac{\Gamma_k^2}{4} - \frac{1}{2}\frac{\dd}{\dd\tau}\Gamma_k \, .
\end{align}

The equations of motion~\eqref{eqn:linear-fourier-modes} take the form
\begin{equation}
  \dot{\bf{y}} = \hat{L}(t){\bf y}
\end{equation}
with
\begin{equation}
  \hat{L}(t+T) = \hat{L}(t) \, ,
\end{equation}
a periodic operator with period $T = 2\pi\nuFreq^{-1}$.
From Floquet theory, solutions take the form
\begin{equation}
  \label{eqn:floquet-mode}
  {\bf y} = e^{\floq \frac{t}{T}}{\bf p}(t) \, ,
\end{equation}
where ${\bf p}(t)$ is a real function satisfying ${\bf p}(t+2T) = {\bf p}(t)$.
The complex constants $\floq$ are known as Floquet exponents, with the solution ${\bf y}$ satisfying~\eqref{eqn:floquet-mode} known as the corresponding Floquet mode.
For~\eqref{eqn:linear-fourier-modes}, ${\bf y} \in \mathbb{R}^2$ and $\tr{\op{L}} = 0$.
It follows that for each choice of parameters $\left(\kBarSym,\nuAmp,\nuBarSym,\wBarSym\right)$ there are two Floquet exponents satisfying $\floq_1 = -\floq_2$, with the two corresponding Floquet modes forming a basis for solutions to the equation.
Furthermore, since $\op{L}$ is real, the Floquet exponents must be either purely real or purely imaginary.
If $\Re(\floq) > 0$, the mode grows exponentially in time, and if the initial condition has a nonzero projection onto this mode, the solution will grow exponentially in time.
The exception to the simple form~\eqref{eqn:floquet-mode} occurs for degenerate Floquet exponents, where an additional power law prefactor is obtained.
However, these degenerate modes generally form a measure zero subset in the space of parameters.

Since we want to study the stability of the proposed false vacuum, we focus on the Floquet exponent with the largest positive real component, which we denote
\begin{equation}\label{eqn:mu-max-def}
  \floqMax \equiv \max\{\Re(\floq_i), i=1,2\} \, .
\end{equation}
With our convention~\eqref{eqn:floquet-mode}, $\floq$ is dimensionless and normalized so that after one period of $\hat{L}$ (i.e. at $t = 2\pi\nuFreq^{-1}$), the corresponding Floquet mode has grown by a factor $e^{\Re(\mu)}$ and acquired a phase $e^{i\Im(\floq)}$.
In order to connect to the either the time-scales of the internal BEC dynamics or the effective relativistic scalar field, we will often find it convenient to work instead with $\wBarSym\floqMax$ or $\nuBarSym^{-1/2}\wBarSym\floqMax$ respectively.

\subsection{Analytic predictions of time-averaged analysis}
The detailed instability structure of the Floquet modes as a function of experimental parameters arises from a complex interplay between the timescale of the external driver $\nuFreq^{-1}$ and the timescale for the undriven dynamics of the modes.
For sufficiently long-wavelengths, the internal time-scale for evolution of the modes will be much longer than $\nuFreq^{-1}$.
In this regime, we therefore expect the time-averaging analysis will provide a suitable approximation to the dynamics, and we can use it to derive analytic predictions for the Floquet exponents.
For shorter wavelengths, this separation of scales will no longer hold, and we expect the time-averaged analysis to fail, an intuition which we will confirm below.
Similarly, for sufficiently large driving amplitudes, we expect that the dynamics of the external driver will become dominant, again invalidating the time-averaging analysis.

We begin with the time-averaged effective Lagrangian for linear perturbations~\eqref{eqn:lag-teff-linear} in the limit $k \to 0$, and define a new field perturbation
\begin{equation}
  \pert{\phiN}_{c} = \left(1 + \cphi\nuBarSym - \nuBarSym\lPar^2\right)^{-1/2}\pert{\phiN} \, ,
\end{equation}
where $\lPar$ and $\nuBarSym$ are defined in~\eqref{eqn:pot-params} and~\eqref{eqn:dimensionless-vars}, respectively.
The corresponding potential for this canonically normalized field is
\begin{equation}
  V_{\rm eff,c} = \cphi\left(1 + \cphi\nuBarSym - \nuBarSym\lPar^2\right)\frac{4\nuZero(\gS-\gC)\rhoPAve}{\hbar^2}\left(1 + \cphi\lPar^2\right)\frac{\pert{\phiN}_c^2}{2} \, .
\end{equation}
The equation of motion for $\delta\phiN_{c}(k=0)$ is
\begin{equation}
  \pert{\ddot{\phiN}}_{c} + V_{\rm eff,c}''\pert{\phiN}_{c} = 0
\end{equation}
so that
\begin{equation}
  \pert{\phiN}_c \sim e^{\pm i t\sqrt{V''_{\rm eff,c}}} \, .
\end{equation}
We immediately read off the prediction for the Floquet exponents:
\begin{equation}
  \label{eqn:floquet-eff-pot-full}
  \nuBarSym^{-1/2}\wBarSym\floq = \pm 4\pi\sqrt{-\cphi}\sqrt{1+\cphi\nuBarSym-\nuBarSym\lPar^2}\sqrt{1+\cphi\lPar^2} \, .
\end{equation}
If we take the limit $\nuBarSym \ll 1$ at fixed $\lPar$ (which here is equivalent to dropping the non-canonical corrections to the kinetic term), we obtain
\begin{equation}
  \label{eqn:floquet-eff-pot-leading}
  \nuBarSym^{-1/2}\wBarSym\floq = \pm 4\pi\sqrt{-\cphi}\sqrt{1+\cphi\lPar^2} \, .
\end{equation}
Below, we compare these predictions of the time-averaging formalism with the full solutions to the linear perturbation equations.
We will see that the time-averaged Hamiltonian analysis accurately reproduces the results of a full Floquet analysis of the long-wavelength modes as long as $\nuBarSym \ll 1$ and $\wBarSym \gtrsim 5$ (in a way that will be made more precise below).
Further, it correctly captures the suppression of the zero-mode instability for increasing $\nuBarSym$.
However, it does not perfectly capture the shift in the stabilizing value of $\lPar$ that occurs as $\nuBarSym$ is increased.
This is not surprising, since the calculation of the effective Hamiltonian proceeds through an expansion in $\nuAmp$ as explained in~\secref{sec:effective-action}.
When $\nuBarSym \sim 1$, we have $\lPar^2 \sim \nuAmp^2$, and thus expect corrections in the intercept which occurs at $\lPar^2 \sim 1$.

\subsection{Floquet Analysis: $\nuAmp \neq 0$}
With the basic formalism in place, we now compute the Floquet exponents for the linear system~\eqref{eqn:linear-fourier-modes}.
Examining the equations of motion, we see that they depend on $\nuBarSym$ and $\kBarSym$ only through the combination
\begin{equation}
  \keffsq \equiv \frac{\kBarSym^2}{4}+\cphi\nuBarSym \, .
\end{equation}
Therefore, we first explore the linear instabilities of the system in terms of the three parameters $\keffsq,\wBarSym$, and $\cphi\nuAmp$.
The results for the maximal Floquet exponent as function of $\keffsq$ and $\cphi\nuAmp$ at various fixed choices of $\wBarSym$ are given in~\figref{fig:floquet-charts-symmetry} and~\figref{fig:floquet-charts-abd-plane}.
In a given experimental realization of the BEC dynamics, the values of $\cphi\nuAmp$ will be fixed, while $\keffsq \in \left[\cphi\nuBarSym,\frac{\tilde{k}_{\rm max}^2}{4}+\cphi\nuBarSym\right]$, where $\tilde{k}_{\rm max}$ is the dimensionless wavenumber above which the GPE description begins to break down.
The inhomogeneous modes properly captured by our analysis would thus live on a vertical line in the Floquet charts (\figref{fig:floquet-charts-abd-plane}).
By increasing $\nuAmp$, the vertical line shifts to the right.
Increasing $\nuBarSym$ either shifts the line down (if ${\cphi\nuBarSym < 0}$) or shifts the line up (if ${\cphi\nuBarSym > 0}$).
If this vertical line intersects a region of instability, we expect the false vacuum decay interpretation of the experiment will be spoiled.

\begin{figure}
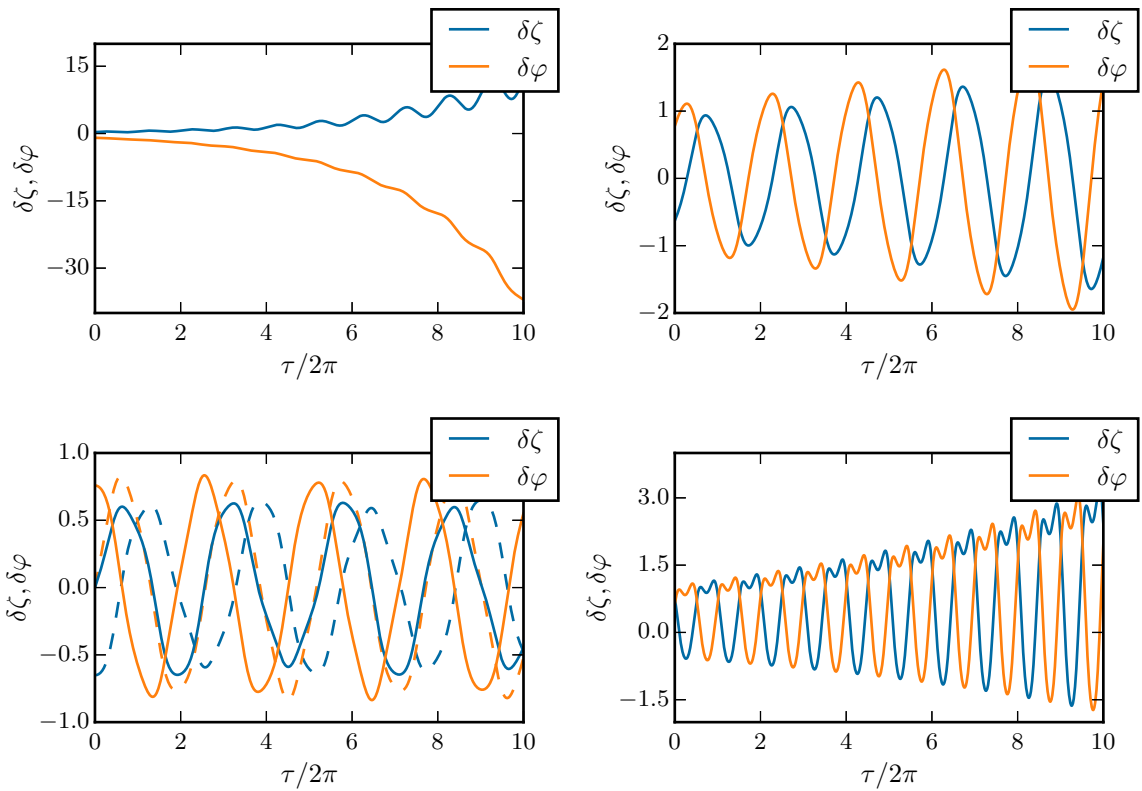

  \includegraphics[width=2.97in]{{{mode-function-kap-0.1-del0.05-w10}}}
  \includegraphics[width=2.97in]{{{mode-function-kscl0.41-del0.05-w10}}}
  \includegraphics[width=2.97in]{{{mode-function-kscl0.3-del0.05-w10}}}
  \includegraphics[width=2.97in]{{{mode-function-kap-0.5-del0.7-w10}}}
  \caption{Some sample Floquet modes for~\eqref{eqn:linear-fourier-modes} illustrating various types of (in)stability.  In each figure, we show the evolution of $\delRatDiff$ (\emph{blue}) and $\pert{\phiN}$ (\emph{orange}).  The mode functions are normalized so $\abs{\delRatDiff}^2 + \abs{\pert{\phiN}}^2 = 1$ initially, and we have only plotted the fastest growing mode (or the mode with positive imaginary Floquet exponent for stable solutions).
    In all cases, we take $\wBarSym=10$.  Solid lines are the real parts of the mode functions, and dashed lines the imaginary part.  For simplicity, we only plot the imaginary components when they are non-vanishing.
    \emph{Top Left}: A tachyonic mode from the broad instability band with $\keffsq=-0.1$ and $2\nuAmp=0.05$.
    \emph{Top Right}: A weak parametric resonance mode in the first band with $2\wBarSym^{-1}\keffsq=0.41$ and $\nuAmp=0.05$. We observe the sign flip of the solution each period of the background due to the $i\pi$ component of the Floquet exponent.
    \emph{Bottom Left}: A stable mode function located within a stability band with $2\wBarSym^{-1}\keffsq=0.3$ and $\nuAmp=0.05$.
    \emph{Bottom Right}: An unstable mode function from an instability triggered by large modulation amplitudes $\nuAmp$, with $\keffsq = -0.5$ and $\nuAmp = 0.7$.}
  \label{fig:mode-funcs}
\end{figure}

Examining~\eqref{eqn:linear-fourier-modes}, one can show the instability charts described in the previous paragraph possess the following symmetry properties for a fixed value of $\wBarSym$:
\begin{enumerate}
\item a reflection symmetry about $\cphi\nuAmp = 0$,
\item a reflection symmetry about the axis $\keffsq = -\frac{1}{2}$, and
\item the chart for $\wBarSym \to -\wBarSym$ is obtained by relecting along the axis $\keffsq=0$.
\end{enumerate}
These properties are illustrated for $\wBarSym=\pm 2$ in~\figref{fig:floquet-charts-symmetry}.
\begin{figure}[ht!]
  \centering
  \includegraphics[width=5.9539in]{{{floquet-ad-symmetry}}}
  \caption{An illustration of the symmetry properties listed in the main text for the specific choice $\wBarSym=2$ (\emph{left}) and $\wBarSym=-2$ (\emph{right}).  As defined in the main text $\keffsq = \frac{\kBarSym^2}{4}+\cphi\nuBarSym$ is an effective wavenumber squared.  The colour scale indicates the value of the maximal Lyapanov exponent $\wBarSym\floqMax$.  White regions where $\floqMax = 0$ indicate the modes are stable.}
  \label{fig:floquet-charts-symmetry}
\end{figure}
Therefore, it is sufficient to plot the instability charts for $\keffsq > -\frac{1}{2}$, $\cphi\nuAmp\geq 0$, and $\wBarSym\geq 0$.
We note that the last condition is the physically meaningful one, as long as $\gS-\gC > 0$.
Some representative examples for various choices of $\wBarSym$ are shown in~\figref{fig:floquet-charts-abd-plane}, where we plot $\floqMax$ for a range of values of $\keffsq$ and $\cphi\nuAmp$.
As is typical for Floquet theory, there exist a series of instability bands emerging from the $\abs{\nuAmp}=0$ axis interspersed with regions of stability.
The behavior of some representative mode functions are shown in~\figref{fig:mode-funcs}.

Examining the various charts in~\figref{fig:floquet-charts-abd-plane}, we see they share a number of common morphological features at each value of $\wBarSym > 0$.
The first important feature is the large instability band centered at $\keffsq=-\frac{1}{2}$ and $\nuAmp=0$ with width $\Delta\keffsq = 1$ at $\nuAmp=0$.
As seen from the top left panel of~\figref{fig:mode-funcs}, the modes within this band experience rapid tachyonic growth with a small superimposed oscillation.
Physically, the modes within this large tachyonic region are well approximated by the time-averaged Hamiltonian analysis presented~\secref{sec:effective-action}.
In addition to this large tachyonic band, there are additional instability bands that appear at
larger values of either $\abs{\keffsq-\frac{1}{2}}$ or $\abs{\cphi\nuAmp}$.
This rich instability structure is not properly captured (even qualitatively) by the time-averaged analysis, since the bare frequencies $\omega_{k,{\rm rel}}$ of these modes are comparable to the driving frequency $\nuFreq$.
The width of these higher order bands vanishes as $\abs{\nuAmp}\to 0$, which is the expected behaviour since $\nuAmp$ parameterizes the strength of the periodic perturbations.
These higher-order bands are separated by regions of stability, where $\floqMax = 0$.
For the range of $\keffsq$ and $\cphi\nuAmp$ values in the plot, the modes within the non-tachyonic instability bands experience weak parametric resonance, undergoing oscillations with an exponentially growing envelope as seen in the upper right panel of~\figref{fig:mode-funcs}.
With our definition of $\keffsq$ these higher order bands move towards the origin and increase in width as $\wBarSym$ is decreased.

Alternatively, if we instead parameterize our equations with $\keffsqt = \wBarSym^{-1}\keffsq$, the locations of the higher order bands are relatively stable as we vary $\wBarSym$.  However, the vertical height of the tachyonic band now shrinks as $\wBarSym$ increases, with a width $2\wBarSym^{-1}$ centered at $\keffsqt = -\wBarSym^{-1}$.
\begin{figure}[ht!]
  \centering
  \includegraphics[width=5.9539in]{{{floquet-charts-ad-plane-panels}}}
  \caption{Floquet charts for~\eqref{eqn:linear-fourier-modes}.  In each panel, we fix the value of $\wBarSym$, and then scan over the effective wavenumber $\keffsq = \frac{\kBarSym^2}{4}+\cphi\nuBarSym$ and $\nuAmp$.  The color scale shows the value of $\wBarSym\floqMax$, thus measuring the growth rate in units of $\frac{\hbar t}{2\pi\rhoPAve(\gS-\gC)}$.  The dashed blue lines correspond to the curve $\keffsq = -2\nuAmp^2$, which is where the stabilization of the zero mode occurs in the leading order time-averaged approximation~\eqref{eqn:floquet-eff-pot-full}. We see deviations from the analytic prediction for $\abs{\nuAmp} \sim \mathcal{O}(1)$, corresponding to $\mathcal{O}(1)$ values of $\abs{\nuBarSym}$.}
  \label{fig:floquet-charts-abd-plane}
\end{figure}

For $\abs{\nuAmp} \ll 1$, the center of the first non-tachyonic instability band follows from energy-momentum conservation for fluctuations produced by a homogeneous source with energy $\hbar\nuFreq$~\cite{Braden:2010wd}.
To leading order in the fluctuation amplitude, the lowest order process in coupling constants produces a pair of modes with equal and opposite wavenumbers.
At this order the modes are produced on shell and the external coupling acts as a source, so energy conservation yields
\begin{equation}
  \hbar\nuFreq = 2\hbar\omega_{k,{\rm rel}} = \sqrt{\left(\frac{\hbar^2k^2}{2m}+2\cphi\nu\right)\left(\frac{\hbar^2k^2}{2m}+2\cphi\nu+2(\gS-\gC)\rhoPAve\right)} \, .
\end{equation}
In terms of dimensionless variables this yields the positive solution
\begin{equation}
  \label{eqn:first-band}
  2\frac{\keffsq}{\wBarSym} = \wBarSym^{-1}\left(\sqrt{1+\frac{\wBarSym^2}{4}}-1\right) \, ,
\end{equation}
from which we see it is convenient to work with the variable $\keffsqt = 2\wBarSym^{-1}\keffsq$ when considering the higher-order instability bands.
In the limits $\wBarSym \gg 1$ and $\wBarSym \ll 1$ we have, respectively
\begin{equation}
  \label{eqn:first-band-limits}
  2\frac{\keffsq}{\wBarSym} \approx \frac{1}{2}-\frac{1}{\wBarSym} \qquad
{\rm and} \qquad
  2\frac{\keffsq}{\wBarSym} \approx \frac{\wBarSym}{8} \, .
\end{equation}
We illustrate this in~\figref{fig:floquet-a-b-plane}, where we see that~\eqref{eqn:first-band} coincides with the center of the first instability band.
Furthermore, for the range $\wBarSym \gg 1$ relevant to analog false vacuum decay, the asymptotic limit provides an accurate approximation.
Similar considerations give the locations of the higher order resonance bands.
\begin{figure}
  \centering
  \includegraphics[width=3.69in]{{{floquet-kap-w-plane}}}
  \caption{Frequency $\wBarSym$ and scaled effective wavenumber $2\wBarSym^{-1}\keffsq$ dependence of $\wBarSym\floqMax$ in the first nontachyonic instability band for the particular choice $\abs{\nuAmp} = 0.05$.  For visibility, the color scale differs from~\figref{fig:floquet-charts-symmetry} and~\figref{fig:floquet-charts-abd-plane}.  We include the analytic prediction for the center of the instability band~\eqref{eqn:first-band} (\emph{yellow dashed}), and the limiting behaviours~\eqref{eqn:first-band-limits} for $\wBarSym \ll 1$ (\emph{magenta solid}) and $\wBarSym \gg 1$ (\emph{black dot-dashed}).  Notice the excellent agreement between the analytic predictions and numerical calculations.}
  \label{fig:floquet-a-b-plane}
\end{figure}

We now estimate the driving frequency $\nuFreq$ required to push this instability band out of the regime of validity of the GPE, assuming for simplicity $\gC = 0$.
Let us denote the smallest non-tachyonic unstable wavenumber $\kBarFloq$.
Inverting~\eqref{eqn:first-band} and using the definition of $g$, we obtain the corresponding driving frequency
\begin{align}
  \nuFreq_{\rm floq} &= 8\pi\left(\frac{\hbar}{m_{\rm pr}a_0^2}\right)\left(\frac{\rhoPAve a_s^3}{Z r^2}\right)\sqrt{\left(\frac{\kBarFloq^2}{2}+1 + 2\cphi\nuBarSym \right)^2 -1} \notag \\
   &\sim 10^{15}\left(\frac{\rhoPAve a_s^3}{Z r^2}\right)\sqrt{\left(\frac{\kBarFloq^2}{2}+1\right)^2 -1}\,\, {\rm Hz} \, .
\end{align}
Here $m_{\rm pr}$ is the proton mass and $a_0$ the Bohr radius.  We have assumed our atoms have nucleon number $Z$ so $m \approx Zm_{\rm p}$ and the $S$-wave scattering length is given by $a_s = r a_0$.

To make a very conservative estimate, we should require that the mode at the center of the instability band $\kFloq$ satisfies $\kFloq = k_{\rm atom}$.
Denoting this frequency by $\nuFreq_{\rm atom}$ we immediately find
\begin{equation}
  \nuFreq_{\rm atom} \approx \frac{4\pi^2}{Z r^2}\frac{\hbar}{m_{\rm pr} a_0^2} \sim 10^{15}\left(\frac{1}{Z r^2}\right)\,\, {\rm Hz} \, .
\end{equation}
We can expect values of $Z r^2 \sim 10^3-10^4$, and therefore the necessary driving frequency is in the GHz. This is somewhat problematic, as hyperfine transitions are associated with energies of order MHz. A less conservative estimate is to require that the instability is at least pushed above the wave number associated with the healing length, which should require somewhat lower modulating frequencies.  By definition $\noDim{k}_{\rm heal} = 1$, so we obtain 
\begin{equation}
  \nuFreq_{\rm heal} \sim 10^{15}\left(\frac{\rhoPAve a_s^3}{Z r^2}\right)\,\, {\rm Hz} \, .
\end{equation}
This is parametrically smaller by the factor $\rhoPAve a_s^3$.
For example, for densities of order $10^{20}\rm{m}^{-3}$ and $S$-wave scattering lengths $a_s \sim 10 a_{0} \sim 1 {\rm nm}$, this yields modulating frequencies a factor $\sim 10^{-3}$ times smaller than $\nuFreq_{\rm atom}$, on the order of 100 kHz to MHz. This is a more consistent modulation for radio frequencies associated with energies of hyperfine transitions. In conclusion, it appears challenging to push the first instability band all the way to $k_{\rm atom}$.

Having obtained the stability charts for the parameters $\keffsq$, $\wBarSym$, and $\cphi\nuAmp$; and briefly commented on their interpretation for the stability of the BECs,
we now quantitatively relate the stability of various modes to the parameters describing the symmetric BECs, namely $\kBarSym,\nuBarSym,\wBarSym$, and $\nuAmp$.

In the context of an analog false vacuum decay experiment, the motivation for introducing the time-dependent oscillations into $\nuT$ is to provide a stabilization mechanism for the unstable fixed point of the background dynamics with $\phiNAve = \pi$.
Therefore, we first demonstrate that this long-wavelength stabilization mechanism operates by studying the dynamics of the $k=0$ mode.
This also allows us to make a quantitative comparison with the predictions of the time-averaged Hamiltonian analysis, from which the emergent false vacua were derived.
\begin{figure}[ht!]
  \includegraphics[width=2.97in]{{{zero-mode-stability-vary-nu}}}
  \includegraphics[width=2.97in]{{{zero-mode-stability-vary-w}}}
  \caption{Stability of the $k=0$ mode as we vary the driving amplitude $\nuAmp$. \emph{Left}: Dependence of $\floqMax$ as we vary $\nuBarSym$ at fixed $\wBarSym = 5$.  Solid lines show the numerical result, and dashed lines the prediction~\eqref{eqn:floquet-eff-pot-full}.  \emph{Right}: Stability for a range of $\wBarSym$ values at fixed $\nuBarSym = 10^{-2}$.  The blue dots show the analytic prediction.}
  \label{fig:floquet-zero-mode}
\end{figure}
\begin{figure}[ht!]
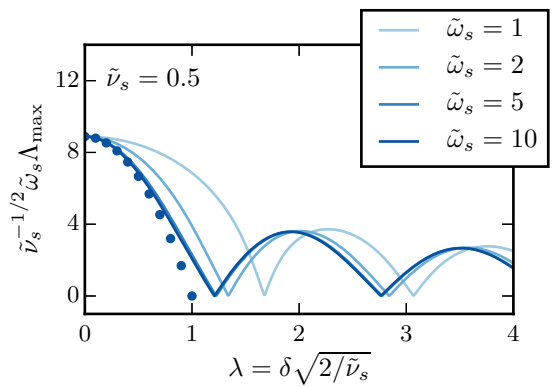

  \centering
  \includegraphics[width=2.97in]{{{zero-mode-stability-vary-w-nu0.5}}}
  \caption{Convergence of the instability in the $k=0$ mode as the driving frequency $\wBarSym$ is increased.  We see convergence to the asymptotic form by $\wBarSym \sim 5$.  The analytic prediction~\eqref{eqn:floquet-eff-pot-full} is shown by the blue dots.}
  \label{fig:zero-mode-w-convergence}
\end{figure}
In~\figref{fig:floquet-zero-mode} we plot $\floqMax(k=0)$ as a function of $\lPar$ for a range of $\nuBarSym$ values with $\wBarSym$ held fixed and vice versa.
For comparison, we also plot the prediction~\eqref{eqn:floquet-eff-pot-full} as dashed lines (left panel) or dots (right panel).
From the left panel, we see for $\nuBarSym \ll 1$ the leading order time-averaging analysis provides an accurate description of the linear instability of the $k=0$ mode.  However, for sufficiently large $\nuBarSym$, we see strong corrections to this relationship emerge.  From the right panel, we see that for $\abs{\nuBarSym} \ll 1$ the shape of the lowest instability band is essentially independent of $\wBarSym$ (when scaled appropriately).  However, as we take $\wBarSym \to 1$ additional instabilities present at larger values of $\lPar$ move towards smaller values of the stabilization parameter $\lPar$.
In~\figref{fig:zero-mode-w-convergence} we further illustrate the convergence properties of the solution of $\wBarSym$ is varied.  We see that even for the large value $\nuBarSym = 0.5$, the zero-mode dynamics has converged for $\wBarSym \gtrsim 5$, indicating that a time-averaged analysis of the long-wavelength modes is sufficient.

\Figref{fig:floquet-zero-mode} demonstrates the stabilisation of the homogeneous degrees of freedom, as predicted by a time-averaged analysis for the zero mode in~\secref{sec:effective-action}.
However, there are additional fluctuations with nontrivial spatial structure whose stability must also be considered.
The internal time-scale for the dynamics of these modes decreases with wavenumber, and for sufficiently short wavelengths the time-averaged analysis is insufficient.
In~\figref{fig:floquet-kscans} we show the stability of the $k\neq 0$ modes as we vary the parameters of the system.
Since our interest in this paper is the case where the homogeneous modes have been stabilized, we choose parameters such that $\lPar = 1.3$ as a representative example.
Even if the zero mode is stabilized, if the range of $\kBarSym$ that are correctly described by the GPE happen to intersect one of the higher order Floquet bands, then the system as a whole will be unstable to a statistically homogeneous amplification of inhomogeneities.
Of course, if the time-scale for these instabilities is much longer than the lifetime of the experiment, then they may be neglected as subleading corrections on the experimental dynamics.
From~\figref{fig:floquet-kscans} we see that bands of higher $k$ modes do indeed experience exponential instabilities.
The location of the bands with respect to $\kBarSym$ has very little sensitivity to $\nuBarSym$ or $\nuAmp$ (i.e.\, $\lPar$).
The primary effect of increasing $\nuBarSym$ is to increase the width of the instability bands and make the higher bands more prominent.
Meanwhile, the primary effect of increasing $\nuAmp$ (or $\lPar$) is to increase the peak strength of the instability bands.
Therefore, the net effect of increasing either of these parameters will be to reduce the time-scale for any unstable Floquet modes present in the system to destabilize the condensate.
However, they cannot be adjusted in such a way as to remove a pre-existing band of unstable modes from the system.

By contrast, in~\figref{fig:floquet-scans-vary-omega}, we see that the primary effect of increasing $\wBarSym$ is to push the instabilities to higher wavenumbers.
In addition, the width of the bands with respect to $\kBarSym$ narrows, while roughly maintaining its peak amplitude.
This second set of properties is robust as long as $\nuBarSym \ll 1$.
We thus identify $\wBarSym$ as the key control parameter to push unstable inhomogeneous modes to shorter wavelengths, where the GPE description eventually fails and we need additional physics to stabilize them to avoid spoiling the false vacuum analog.
This is, of course, entirely expected since $\wBarSym^{-1}$ measures the ratio of timescales associated with the internal scattering of the individual atoms to that of the external modulation.
By increasing $\wBarSym$, these scattering interactions must occur on much shorter timescales to respond to $\wBarSym$, corresponding to much larger wavenumbers.

\begin{figure}[ht!]
  \includegraphics[width=2.97in]{{{k-stability-symmetric-vary-lam}}}
  \includegraphics[width=2.97in]{{{k-stability-symmetric-vary-nu}}}
  \caption{Stability charts as a function of (dimensionless) wavenumber $\kBarSym$.  \emph{Left}: We vary the driving amplitude $\lPar$ fixed while $\wBarSym$ and $\nuBarSym$ are held constant.   \emph{Right}: We vary the amplitude of the constant inter-condensate conversion $\nuBarSym$ while holding $\wBarSym$ and $\lPar$ fixed.  Here we have scaled the Floquet exponents by $\nuBarSym^{-1/2}\wBarSym$ as appropriate for the dynamics of the effective scalar field. For $\nuBarSym \ll 1$, the location of the instability band (in $\kBarSym$) is essentially independent of both $\lPar$ and $\nuBarSym$.  In these units, increasing $\lPar$ both the stabilizes the zero mode and increases the strength of the higher order Floquet bands. Meanwhile, increasing $\nuBarSym$ primarily increases the width of the instability band, while slightly decreasing its amplitude.}
  \label{fig:floquet-kscans}
\end{figure}

\begin{figure}
  \centering
  \includegraphics[width=2.97in]{{{k-stability-symmetric-vary-w}}}
  \caption{A stability chart as in~\figref{fig:floquet-kscans} except now varying the driving frequency $\wBarSym$ at fixed $\nuBarSym$ and $\lPar$ (or equivalently $\nuAmp$).  Unlike the variations of $\lPar$ and $\nuBarSym$ illustrated in~\figref{fig:floquet-kscans}, we see that the value of $\wBarSym$ has a dramatic effect on the location of the bands.  As expected, increasing $\wBarSym$ pushes the instability to larger values of $\kBarSym$.}
  \label{fig:floquet-scans-vary-omega}
\end{figure}

\section{Conclusions and Future Work}\label{sec:conclusion}
We undertook a thorough analysis of the analog relativistic false vacuum decay proposal of Fialko et al~\cite{Fialko:2014xba,Fialko:2016ggg}.
If experimentally feasible, this provides the exciting possibility of testing strongly nonlinear and nonperturbative quantum field dynamics in a fully time-dependent setting.
The specific phenomena under consideration, relativistic false vacuum decay, has direct application to multiverse models of cosmology.
Due to the exponential complexity of quantum field theory, no tools exist to treat the decay problem exactly.
Furthermore, extant approximations do not rigorously incorporate the time-dependent nature of the decay, nor can they be used to explore correlations between tunnelling events.
When applied to cosmology, these correlations can significantly influence the distribution of fundamental constants seen by observers in the multiverse.
Furthermore, they can have important implications for the probability of observing signatures of the multiverse through bubble collisions.
Reaching a deeper understanding of these issues can thus shed light on some of the most fundamental issues in modern cosmology.

In order to assess the feasibility of the proposal, we first carefully derived the relationship between the theoretical description of the BEC and the emergent relativistic scalar field with a false vacuum potential.
We cataloged the various corrections to the simple interpretation as an effective canonical scalar field, and indicated the required conditions for these corrections to be small.
Since the only aspect of the BEC physics used in the derivation was the applicability of the GPE, this illustrates the generality of the mechanism beyond any specific experimental proposal.
We then explored the linear stability properties of the proposed false vacua using Floquet theory. The introduction of external modulations in the interspecies conversion rate (which was the key ingredient to generate effective false vacuum minima) simultaneously induces new bands of instabilities in the shorter wavelengths.  Furthermore, for a sufficiently strong modulation amplitude, the long-wavelength modes can again be destabilized.

The key control parameter determining the wavelengths of the newly destabilized modes is the frequency of the modulations in the interspecies conversion rate $\wBarSym$.
We demonstrated that the corresponding instability wavenumber scales as $\sqrt{\wBarSym}$ for the relevant regime $\wBarSym \gg 1$.
Since the GPE description does not hold to arbitrarily short wavelengths, it is possible to push the higher order Floquet instabilities out of the regime of validity of the coupled GPEs studied here.  Physically, we expect the discreteness of the particles will induce viscosity and stochasticity into the equations at short wavelengths, which will act to damp the exponential growth of the modes.
In the pessimistic scenario where modes up to $k_{\rm atom}$ obey the GPE, driving frequencies of order of tens of GHz are required.
In the extremely optimistic scenario where modes beyond the healing length are damped, frequencies in the 100 kHz range are instead needed.
However, one must also ensure that the increased modulation frequency does not allow additional modes of the system to be excited, such as the radial eigenmodes of a ring trap.
For the particular setup described in~\cite{Fialko:2016ggg} based on radio-frequency fields, ensuring the required frequency is below the MHz scale requires that the linear dynamics of modes a factor of a few beyond the healing length must be modified from the GPE description.
Whether this condition holds in reality requires further investigation.

In order to illustrate the key dynamics with minimal technical complications, we made two key assumptions in our analysis: that the background condensate is spatially homogeneous and that the two particle species are symmetric (with equal masses and interspecies scattering rates).
On technical grounds, the first assumption allows for a diagonalization of the fluctuation modes through the Fourier transform, explicitly decoupling the equations of motion and transforming a PDE into a collection of uncoupled ODEs.
Physically, one could imagine the condensates are trapped in a ring with strong trapping potentials in only two directions.
In this case the effective one-dimensional dynamics should be well approximated by the spatially homogeneous and periodic setup.
However, one needs to be careful not to excite the radial eigenmodes of the system through modulations in the transition rate.
Alternatively, accounting for the inhomogeneity of the background induced by the external trap, one can imagine scenarios where there is a  narrow boundary layer near the edges of the trap, and then a nearly homogeneous plateau in the middle.  Our analysis should then account for the dynamics of perturbations with wavelengths much smaller than the width of the plateau.
An interesting extension is to include the dynamical effects of the trapping potentials for the condensate, which will introduce inhomogeneity into the background.
Assuming there is still a false vacuum interpretation in this case, the fluctuations on this inhomogeneous background can then be studied using the methods developed in~\cite{Braden:2014cra}.

The symmetric assumption (i.e. $g_{11}=g_{22}$ and $m_1=m_2$) along with $\rhoNAve = 0$ was primarily adopted for technical simplicity and to create the most direct connection between the cold atom system and the relativistic false vacuum decay system.
Relaxing these conditions generally leads to a mixing between the total and relative phonons, so that the effective field theory description describes two coupled relativistic fields.
Further, the sound speeds between the modes need not be the same, so the relativistic interpretation can be spoiled.
From the viewpoint of linear stability, the coupling between the modes can lead to the emergence of a second unstable mode, modify the stabilization condition for the zero mode, and shift the value of $\kBarSym$ at which the first non-tachyonic instability band emerges.
The rich resulting dynamics will be explored in a future publication.

As well as these extensions, an interesting possibility is to consider the interactions between many condensates, rather than just two.
This could allow for an experimental realization of a complex potential landscape, analogous to those believed to descend from low-energy string theory constructions and that form a fundamental ingredient in many current multiverse paradigms for cosmology.
We do not expect the linear stability analysis to differ dramatically with the addition of additional condensates.
On the contrary, the nonlinear dynamics resulting from the decay of the false vacuum may display considerably more variability with the introduction of additional decay and entropy production channels.
Since this paper focuses solely on the linear stability analysis of the ``UV complete'' theory of the Bose-Condensates, we leave these interesting possibilities to future work.

\acknowledgments
We thank P. Drummond for introducing us to his fascinating work, and Sebastian Erne for stimulating discussions and feedback on the manuscript. MCJ is supported by the National Science and Engineering Research Council through a Discovery grant. JB and HVP are supported the European Research Council under the European Community's Seventh Framework Programme (FP7/2007-2013) / ERC grant agreement no 306478-CosmicDawn. This work was partially enabled by funding from the University College London (UCL) Cosmoparticle Initiative. This work was supported in part by National Science Foundation Grant No. PHYS-1066293 and the hospitality of the Aspen Center for Physics. This research was supported in part by Perimeter Institute for Theoretical Physics. Research at Perimeter Institute is supported by the Government of Canada through the Department of Innovation, Science and Economic Development Canada and by the Province of Ontario through the Ministry of Research, Innovation and Science. SW acknowledges financial support provided under the Royal Society University Research Fellow (UF120112), the Nottingham Advanced Research Fellow (A2RHS2), the Royal Society Project (RG130377) grants, and the EPSRC Project Grant (EP/P00637X/1).

\appendix

\section{Effective Action for Condensate Phases}\label{app:effective-action}
In this appendix we derive the effective action for the total and relative phases under the assumption of symmetric condensates (i.e.\ $m_1=m_2$, $g_{11}=g_{22}$) and a homogeneous background around which to expand our fluctuations.
Furthermore, we assume that we can fix the dynamical background variable $\rhoNAve = 0$.
This is a common approximation in the literature, but we note that a full consideration of the field dynamics may break the assumption $\rhoNAve = 0$.
We explicitly indicate the required approximations to arrive at an action describing a relativistic field with canonical kinetic term, as well as provide some arguments for why these approximations may hold.
A fully rigorous justification requires tracking the dynamics of the homogeneous background, including the asymmetry between the background condensate densities $\rhoNAve$.
However, even with the restriction to symmetric condensates, the technical details are already rather cumbersome, so we leave an investigation of the effects of introducing asymmetries between the condensates to future work.

We first split the density fields into a homogeneous background plus an inhomogeneous fluctuation:
\begin{subequations}
\begin{align}
  \rhoP({\bf x},t) &= \rhoPAve(t) + \pert{\rhoP}({\bf x},t)\, , \\
  \rhoN({\bf x},t) &= \rhoNAve(t) + \pert{\rhoN}({\bf x},t)\, .
\end{align}
\end{subequations}
As written, this does not uniquely define the perturbations due to ambiguity in the $k=0$ mode amplitude.
We fix this ambiguity by forcing $\langle\pert{\rhoP}\rangle = \langle\pert{\rhoN}\rangle = 0$, where $\langle\cdot\rangle$ represents a spatial average.
However, we note that in some cases it is convenient to demand that $\rhoPAve$ and $\rhoNAve$ (along with the corresponding averages for the angles) follow the homogeneous background equations of motion as presented in~\secref{sec:homogeneous-bg}.

The first step is to expand the Lagrangian density
to quadratic order in the density perturbations $\pert{\rhoP}$ and $\pert{\rhoN}$.
Since we are interested in the nonlinear dynamics of the phase variables, we \emph{do not} expand them in fluctuations.
Denoting the $i$th order of this expansion by $\lagDdel[i]$, a straightforward calculation gives
\begin{subequations}
\begin{align}
  \lagDdel[0] &= -\hbar\rhoPAve\dot{\phiP} - \frac{\hbar^2\rhoPAve}{4m}(\nabla\phiP)^2 - \frac{\hbar^2\rhoPAve}{4m}(\nabla\phiN)^2 + 2\nuT\rhoPAve\cos\phiN - (\gS+\gC)\rhoPAve^2 + 2\chemP \rhoPAve \\
  \lagDdel[1] &= \left(\begin{array}{c}
                -\hbar\dot{\phiP} - 2(\gS+\gC)\rhoPAve + 2\nu\cos\phiN - \frac{\hbar^2}{4m}(\nabla\phiP)^2 - \frac{\hbar^2}{4m}(\nabla\phiN)^2 + 2\chemP \\ 
                -\hbar\dot{\phiN} - \frac{\hbar^2}{2m}\nabla\phiP\cdot\nabla\phiN
                 \end{array}\right)^T
           \left(\begin{array}{c} \pert{\rhoP} \\ \pert{\rhoN} \end{array}\right) \\
  \lagDdel[2] &= -\frac{1}{2}\left(\begin{array}{c} \pert{\rhoP} \\ \pert{\rhoN} \end{array}\right)^T
             \left(\begin{array}{cc} 2(\gS+\gC) +  \frac{\hbar^2\nabla^T\cdot\nabla}{2\rhoPAve m} & 0 \\ 
                       0 & 2(\gS-\gC) + 2\frac{\nu}{\rhoPAve}\cos\phiN + \frac{\hbar^2\nabla^T\cdot\nabla}{2\rhoPAve m} \end{array}\right) 
             \left(\begin{array}{c} \pert{\rhoP} \\ \pert{\rhoN} \end{array}\right) \, .
\end{align}
\end{subequations}
At the next order in the expansion, we obtain terms schematically of the form
\begin{equation}
  \sim\frac{\delta\rho_i}{\rhoPAve}\frac{\hbar^2(\nabla\rho_i)^2}{\rhoPAve m} \qquad \mathrm{and} \qquad \sim \frac{\nu}{\rhoPAve}\frac{\pert{\rho}}{\rhoPAve}\pert{\rho}^2\cos\phiN \, ,
\end{equation}
where $i=1,2$ labels the individual condensate species and $\pert{\rho}$ represents fluctuations in either condensate.
Smallness of $\pert{\rhoN}$ and $\pert{\rhoP}$ thus ensures that we are keeping the leading terms involving derivatives of the density perturbations.
In cases where $\rhoPAve\pert{\phi} \lesssim \pert{\rho}$, (with $\pert{\phi}$ representing fluctuations in any of the phase variables) consistency also requires us to drop the terms involving two derivatives of the phase variables in $\lagDdel[1]$.
An example of such a state is one where the condensate fluctuations $\pert{\psi}_i$ are complex Gaussian random fields.  For general states this condition need not hold. 
However, comparing to the $\mathcal{O}(\nabla^2)$ terms in $\lagDdel[0]$, we see that the contributions will be suppressed by the smallness of the density perturbations.
In the interests of simplicity, we will therefore neglect the $\mathcal{O}(\nabla^2)$ contributions to $\lagDdel[1]$.

Before proceeding, we also divide the total phase into a homogeneous background and perturbation
\begin{equation}
  \phiP({\bf x},t) = \phiPAve(t) + \pert{\phiP}({\bf x},t) \, .
\end{equation}
As with the density perturbations, it may be convenient to define the background $\phiPAve$ via either removal of the tadpole $\langle\pert{\phiP}\rangle = 0$, or to define $\phiPAve(t)$ to be a solution of the homogeneous background equations.
We now introduce the quantity
\begin{equation}
  \eomPhiT[\phiPAve,\phiNAve,\rhoPAve] \equiv -\hbar\dot{\phiPAve} - 2(\gS+\gC)\rhoPAve + 2\nuT\cos\phiNAve + 2\chemP
\end{equation}
which is the homogeneous background equation of motion for $\phiPAve$ when $\rhoNAve = 0$.

By integrating the Gaussian density perturbations $\pert{\rhoP}$ and $\pert{\rhoN}$, we obtain an effective Lagrangian for the phase variables
\begin{align}
  \label{eqn:lag-eff-quad-rho}
  \mathcal{L}_{\rm eff}^{\phiP,\phiN} &= -\hbar\rhoPAve\dot{\phiPAve} - (\gS+\gC)\rhoPAve^2 + 2\chemP\rhoPAve + 2\nuT\rhoPAve\cos\phiNAve + \frac{1}{4(\gS+\gC)}\eomPhiT^2 \notag \\
    &+ \frac{\hbar^2}{4(\gS+\gC)}\pert{\dot{\phiP}}\left(1-\frac{\hbar^2\nabla^2}{4m(\gS+\gC)\rhoPAve}\right)^{-1}\pert{\dot{\phiP}} - \frac{\hbar^2\rhoPAve}{4m}(\nabla\pert{\phiP})^2 -\hbar\rhoPAve\pert{\dot{\phiP}} \notag \\
    &+ \frac{\hbar^2}{4(\gS-\gC)}\dot{\phiN}\left(1+\frac{\nuT\cos\phiN}{(\gS-\gC)\rhoPAve} - \frac{\hbar^2\nabla^2}{4m(\gS-\gC)\rhoPAve}\right)^{-1}\dot{\phiN} -\frac{\hbar^2\rhoPAve}{4m}(\nabla\phiN)^2 + 2\nuT\rhoPAve\left(\cos\phiN-\cos\phiNAve\right) \notag \\
  &+ \frac{1}{2V}\ln\mathrm{det}\left(1+\frac{\nuT\cos\phiN}{(\gS-\gC)\rhoPAve} - \frac{\hbar^2\nabla^2}{4m(\gS-\gC)\rhoPAve}\right) \notag \\
  &+ \frac{\eomPhiT}{2(\gS+\gC)}\left(-\hbar\pert{\dot{\phiP}} + 2\nuT(\cos\phiN-\cos\phiNAve)\right) \notag \\
  &+ \frac{\nuT^2}{(\gS+\gC)}\left(\cos\phiN-\cos\phiNAve\right)\left(1-\frac{\hbar^2\nabla^2}{4m\rhoPAve(\gS+\gC)}\right)^{-1}\left(\cos\phiN-\cos\phiNAve\right) \notag \\
  &-\hbar\frac{\nuT}{(\gS+\gC)}\left(\cos\phiN-\cos\phiNAve\right)\left(1-\frac{\hbar^2\nabla^2}{4m\rhoPAve(\gS+\gC)}\right)^{-1}\pert{\dot{\phiP}} \, .
\end{align}
Here we have ignored spatial dependence of the functional determinant to put it back inside the integral.
Since we will not consider the effects of this term in the remainder of the paper, this approximation has no impact on our results.
We also assume we are working in an experimental regime where ${\nuT/(\gS-\gC)\rhoPAve \ll 1}$, and restrict ourselves to the long-wavelength modes with ${\hbar^2k^2/4(\gS\pm\gC)m\rhoPAve \ll 1}$.
The effective action~\eqref{eqn:lag-eff-quad-rho} then simplifies
\begin{align}
  \label{eqn:lag-eff-quad-rho-klim}
  \mathcal{L}_{\rm eff}^{\phiP,\phiN} &= -\hbar\rhoPAve\dot{\phiPAve} - (\gS+\gC)\rhoPAve^2 + 2\chemP\rhoPAve + 2\nuT\rhoPAve\cos\phiNAve + \frac{1}{4(\gS+\gC)}\eomPhiT^2 \notag \\
    &+ \frac{\hbar^2}{4(\gS+\gC)}\pert{\dot{\phiP}}^2 - \frac{\hbar^2\rhoPAve}{4m}(\nabla\pert{\phiP})^2 - \hbar\rhoPAve\pert{\dot{\phiP}} \notag \\
  &+ \frac{\hbar^2}{4(\gS-\gC)}\dot{\phiN}^2 -\frac{\hbar^2\rhoPAve}{4m}(\nabla\phiN)^2 + 2\nuT\rhoPAve\left(\cos\phiN-\cos\phiNAve\right) \\
  &+ \frac{2\eomPhiT}{4(\gS+\gC)}\left(-\hbar\pert{\dot{\phiP}} + 2\nuT(\cos\phiN-\cos\phiNAve)\right) + \frac{\nuT^2}{(\gS+\gC)}\left(\cos\phiN-\cos\phiNAve\right)^2 \notag \\
  & -\frac{\nuT}{(\gS+\gC)}\hbar\pert{\dot{\phiP}}\left(\cos\phiN-\cos\phiNAve\right) \, . \notag
\end{align}
Finally, if we ignore the backreaction of the fluctuations on the homogeneous dynamics, then the barred quantities will follow the equations of motion for a homogeneous background as presented in~\secref{sec:homogeneous-bg}.
From this, we see that $\eomPhiT = 0$ on background solutions with $\rhoNAve = 0$, so we are left with
\begin{align}
  \label{eqn:lag-eff-quad-two-phases}
  \mathcal{L}_{\rm eff}^{\phiP,\phiN} &= -\hbar\rhoPAve\dot{\phiPAve} - (\gS+\gC)\rhoPAve^2 + 2\chemP\rhoPAve + 2\nuT\rhoPAve\cos\phiNAve \notag \\
    &+ \frac{\hbar^2}{4(\gS+\gC)}\pert{\dot{\phiP}}^2 - \frac{\hbar^2\rhoPAve}{4m}(\nabla\pert{\phiP})^2 -\frac{\nuT}{(\gS+\gC)}\hbar\pert{\dot{\phiP}}\left(\cos\phiN-\cos\phiNAve\right) \notag \\
  &+ \frac{\hbar^2}{4(\gS-\gC)}\dot{\phiN}^2 -\frac{\hbar^2\rhoPAve}{4m}(\nabla\phiN)^2 + 2\nuT\rhoPAve\left(\cos\phiN-\cos\phiNAve\right) \\
  &+ \frac{\nuT^2}{(\gS+\gC)}\left(\cos\phiN-\cos\phiNAve\right)^2 \, . \notag
\end{align}
Since we have assumed the fluctuations do not backreact or rescatter off the background, the first line of~\eqref{eqn:lag-eff-quad-two-phases} acts as an irrelevant constant for the dynamics of the perturbations.

Under the sequence of approximations given above, we end up with an effective Lagrangian for the phase variables that is quadratic in the total phase perturbations $\pert{\phiP}$.
Therefore, we can also integrate out the perturbations $\pert{\phiP}$ to obtain an effective action for the relative phase variable $\phiN$ alone.
This is most easily done by going to Euclidean time $\tau_E = it$, and expanding $\pert{\phiP}$ is Matsubara modes.
We will neglect these additional environmental contributions here, but we mention in passing 
a subtlety that appears if the Euclidean time average of $\partial_{\tau_E}\pert{\phiP}$ is non-vanishing.
In this case, the zero mode of $\pert{\phiP}$ is not representable as part of the $\tau_E$-periodic Matsubara sum and must be treated separately, generating additional environmental contributions.
Truncating the $\phiP$ dynamics and working to leading order in $\nuT/(\gS+\gC)\rhoPAve$, we obtain the effective sine-Gordon Lagrangian~\eqref{eqn:lag-eff-sine-gordon} presented in~\secref{sec:effective-action}.

\subsection{Effective Action for Linear Perturbations}
To directly connect to the linear fluctuation analysis presented in this paper, it is also of interest to obtain the effective Lagrangian for linear perturbations to the phase variables around our fiducial background false vacuum states.
We demonstrate that under suitable approximations, the effective action obtained this way produces equivalent results to direct manipulation of the equations of motion.
As above, due to the complexity of the resulting equations, we will restrict ourselves to the case of symmetric condensates here.
Furthermore, we assume the background is fixed, with $\cos\phiNAve = \pm 1 \equiv \sigma$.

Since we now wish to concentrate on the equations for linear perturbations around a homogeneous background, we write
\begin{subequations}
\begin{align}
  \phiP({\bf x},t) &= \phiPAve(t) + \pert{\phiP}({\bf x},t) \\
  \phiN({\bf x},t) &= \phiNAve(t) + \pert{\phiN}({\bf x},t)
\end{align}
\end{subequations}
with $\int \dd^dx \pert{\phiN} = 0 = \int \dd^dx \pert{\phiP}$.
Expanding the Lagrangian density to quadratic order in $\pert{\phiP}$ and $\pert{\phiN}$ (in addition to $\pert{\rhoP}$ and $\pert{\rhoN}$), we see that the $\mathcal{O}(\nabla^2)$ contributions to $\lagDdel[1]$ disappear, and that $\lagDdel[2]$ becomes independent of the phase fluctuations.
Therefore, we can begin directly with~\eqref{eqn:lag-eff-quad-rho}, except with the fluctuation determinant on the fourth line set to a constant.
Further, we perturb around a homogeneous background solution so $\eomPhiT = 0$.
Expanding~\eqref{eqn:lag-eff-quad-rho} to quadratic order in $\pert{\phiP}$ and $\pert{\phiN}$ we see the dynamics of the total and relative phonon modes decouple.
We are left with the following pair of effective Lagrangian densities
\begin{equation}\label{eqn:lag-eff-linear-tot}
  \frac{2(\gS+\gC)\rhoPAve}{\hbar^2\rhoPAve}\mathcal{L}_{\rm lin}^{\phiP} = \frac{1}{2}\pert{\dot{\phiP}}\left(1+\frac{\hbar^2\nabla^T\cdot\nabla}{4m\rhoPAve(\gS+\gC)}\right)^{-1}\pert{\dot{\phiP}} - \frac{\rhoPAve(\gS+\gC)}{m}\frac{(\nabla\pert\phiP)^2}{2} 
\end{equation}
and
\begin{align}\label{eqn:lag-eff-linear-rel}
  \frac{2(\gS-\gC)\rhoPAve}{\hbar^2\rhoPAve}\mathcal{L}_{\rm lin}^{\phiN} &= \frac{1}{2}\pert{\dot{\phiN}}\left(1 + \frac{\nuT\cos\phiNAve}{(\gS-\gC)\rhoPAve} + \frac{\hbar^2\nabla^T\cdot\nabla}{4m\rhoPAve(\gS-\gC)} \right)^{-1}\pert{\dot{\phiN}} \notag \\
  &- \frac{\rhoPAve(\gS-\gC)}{m}\frac{(\nabla\pert{\phiN})^2}{2} - \frac{2\nuT\rhoPAve(\gS-\gC)}{\hbar^2}\cos\phiNAve\frac{\pert{\phiN}^2}{2} \, .
\end{align}
In~\eqref{eqn:lag-eff-linear-tot} we have dropped a total time derivative.
Using the Euler-Lagrange equations on each of these effective Lagrangians, we obtain
\begin{equation}\label{eqn:eom-tot-phase-lag}
  \pert{\ddot{\phiP}} - \frac{\rhoPAve(\gS+\gC)}{m}\left(1-\frac{\hbar^2\nabla^2}{4m\rhoPAve(\gS+\gC)}\right)\nabla^2\pert{\phiP} = 0
\end{equation}
and
\begin{equation}\label{eqn:eom-rel-phase-lag}
  \frac{\dd}{\dd t}\left(\frac{1}{1 + \frac{\nuT\cos\phiNAve}{(\gS-\gC)\rhoPAve} - \frac{\hbar^2\nabla^2}{4m\rhoPAve(\gS-\gC)}}\pert{\dot{\phiN}}\right) - \frac{\rhoPAve(\gS-\gC)}{m}\nabla^2\pert{\phiN} - \frac{2\nuT\rhoPAve(\gS-\gC)}{\hbar^2}\pert{\phiN} = 0
\end{equation}
in agreement with~\eqref{eqn:wave-eqn-tot-phase-sym} and~\eqref{eqn:wave-eqn-rel-phase-sym} respectively.

\section{Linear Perturbation Equations}\label{app:linear}
In this section we briefly outline the derivation of the perturbation equations for linear fluctuations around homogeneous background solutions of the GPE.
We write $\rho_i = \rhoM_i +\rhoPert_i = \rhoM_i\left(1+\rhoNorm_i\right)$ and ${\phi_i = \phaseM_i + \phasePert_i}$.
For the case of a two condensate system with $g_{11}=g_{22}=g$ and $m_{11}=m_{22}=m$, we have to linear order in the perturbations:
\begin{subequations}\label{eqn:pert-eqns-linbasis}
\begin{align}
  \hbar\frac{\dd\rhoPert_1}{\dd t} &= -\frac{\hbar^2}{m}\rhoM_1\nabla^2\phasePert_1 - \nuT\sqrt{\rhoM_1\rhoM_2}\sin\phiNAve(\rhoNorm_1+\rhoNorm_2) -2\nuT\sqrt{\rhoM_1\rhoM_2}\cos\phiNAve(\phasePert_2-\phasePert_1) \\
  \hbar\frac{\dd\rhoPert_2}{\dd t} &= -\frac{\hbar^2}{m}\rhoM_2\nabla^2\phasePert_2 + \nuT\sqrt{\rhoM_1\rhoM_2}\sin\phiNAve(\rhoNorm_1+\rhoNorm_2) + 2\nuT\sqrt{\rhoM_1\rhoM_2}\cos\phiNAve(\phasePert_2-\phasePert_1) \\
  \hbar\frac{\dd\phasePert_1}{\dd t} &= \frac{\hbar^2}{4m\rhoM_1}\nabla^2\rhoPert_1 - \gS\rhoPert_1 - \gC\rhoPert_2 + \nuT\sqrt{\frac{\rhoM_2}{\rhoM_1}}\left(\cos\phiNAve\frac{(\rhoNorm_2-\rhoNorm_1)}{2} - \sin\phiNAve(\phasePert_2-\phasePert_1)\right)\\\
  \hbar\frac{\dd\phasePert_2}{\dd t} &= \frac{\hbar^2}{4m\rhoM_2}\nabla^2\rhoPert_2 - \gS\rhoPert_2 - \gC\rhoPert_1 - \nuT\sqrt{\frac{\rhoM_1}{\rhoM_2}}\left(\cos\phiNAve\frac{(\rhoNorm_2-\rhoNorm_1)}{2} + \sin\phiNAve(\phasePert_2-\phasePert_1)\right) \, .
\end{align}
\end{subequations}

We now convert to our preferred variables, the relative and total phonon modes.
To simplify some expressions, we recall the dimensionless fractional difference in the background densities defined in~\secref{sec:homogeneous-bg},
\begin{equation}
  \drhoRat = \frac{\rhoNAve}{\rhoPAve} \, .
\end{equation}
After some straightforward manipulations we obtain
\begin{subequations}\label{eqn:lin-eqns-bg-tdep}
\begin{align}
  \hbar\frac{\dd\pert{\rhoP}}{\dd t} &= -\frac{\hbar^2\rhoPAve}{2m}\left(\nabla^2\pert{\phiP} + \drhoRat\nabla^2\pert{\phiN}\right) \, ,\\
  \hbar\frac{\dd\pert{\phiP}}{\dd t} &= \frac{\hbar^2}{2m\rhoPAve}\left(1-\drhoRat^2\right)^{-1}\left(\nabla^2\pert{\rhoP} - \drhoRat\nabla^2\pert{\rhoN}\right) - 2(\gS+\gC)\pert{\rhoP} \notag \\
  &+ 2\frac{\nuT}{\rhoPAve}\cos\phiNAve\frac{\drhoRat}{\left(1-\drhoRat^2\right)^{3/2}}\left(\pert{\rhoN}-\drhoRat\pert{\rhoP}\right)
  - \frac{2\nuT}{\sqrt{1-\drhoRat^2}}\sin\phiNAve\pert{\phiN} \, ,\\
  \hbar\frac{\dd\pert{\rhoN}}{\dd t} &= -\frac{\hbar^2\rhoPAve}{2m}\left(\nabla^2\pert{\phiN} + \drhoRat\nabla^2\pert{\phiP}\right) \notag \\
  &+ 2\nuT\rhoPAve\sqrt{1-\drhoRat^2}\cos\phiNAve\pert\phiN + \frac{2\nuT}{\sqrt{1-\drhoRat^2}}\sin\phiNAve\left(\pert{\rhoP}-\drhoRat\pert{\rhoN}\right) \, ,\\
  \hbar\frac{\dd\pert{\phiN}}{\dd t} &= \frac{\hbar^2}{2m\rhoPAve}\left(1-\drhoRat^2\right)^{-1}\left(\nabla^2\pert{\rhoN} - \drhoRat\nabla^2\pert{\rhoP}\right) - 2(\gS-\gC)\pert{\rhoN} \notag \\
  &- 2\frac{\nuT}{\rhoPAve}\left(1-\drhoRat^2\right)^{-3/2}\cos\phiNAve\left(\pert{\rhoN}-\drhoRat\pert{\rhoP}\right) + 2\nuT\sin\phiNAve\frac{\drhoRat}{\sqrt{1-\drhoRat^2}}\pert{\phiN} \, .
\end{align}
\end{subequations}
At this point, these equations are valid for linear perturbations around an arbitrary homogeneous (but possibly time-dependent) background.
In particular, if the background is allowed to evolved dynamically, these equations can be used to study self-resonant production of fluctuations, which may allow for an analog experiment to simulate some aspects of the post-inflationary stage of preheating in the sine-Gordon model~\cite{Greene:1998pb}.

As shown in the main text, for the proposed false vacuum the interesting perturbation dynamics occur exclusively in the relative phonons.  Therefore, it is convenient to introduce the dimensionless units
\begin{equation}
  \tRel \equiv \frac{(\gS-\gC)\rhoPAve}{\hbar}\, t \qquad \xRel \equiv \sqrt{\frac{m}{(\gS-\gC)\rhoPAve}}\frac{(\gS-\gC)\rhoPAve}{\hbar}\, x  \, .
\end{equation}
For general nonlinear dynamics and numerical work, it can be convenient to instead introduce
\begin{equation}
  \tG \equiv \frac{\gS\rhoPAve}{\hbar}\, t \qquad \xG \equiv \sqrt{\frac{m}{\gS\rhoPAve}}\frac{\gS\rhoPAve}{\hbar}\, x \, .
\end{equation}

\bibliography{linear-stability}
\bibliographystyle{JHEP}

\end{document}